\documentclass[prl,amsmath,amssymb,aps,superscriptaddress,floatfix,reprint,notitlepage]{revtex4-1}
\usepackage{natbib}
\usepackage{graphicx}
\usepackage{color}
\usepackage{dcolumn}
\usepackage{bm}
\usepackage[utf8]{inputenc}
\usepackage{amssymb}
\usepackage{color,amsmath}
\usepackage{mathtools,ulem}
\usepackage{soul,xcolor} 
\setstcolor{red} 
\usepackage{epstopdf}
\usepackage{scalerel}
\usepackage{hyperref}
\hypersetup{
 pdfnewwindow=true, colorlinks=true,
 linkcolor=blue, anchorcolor=blue,
 citecolor=blue, filecolor=blue,
 menucolor=blue, urlcolor=blue}

\newcommand{\change}[2]{{\color{red}\sout{#1}}{\color{blue}#2}}
\newcommand{\auc}{A_\mathrm{u.c.}}
\newcommand{\sub}[1]{\scaleto{\text{#1}}{4pt}}

\begin{document}

\title{Electrical switching of magnetic order in an orbital Chern insulator}
\author{H. Polshyn} 
\affiliation{Department of Physics, University of California, Santa Barbara, CA 93106}
\author{J. Zhu} 
\affiliation{Department of Physics, University of Texas, Austin, TX 78712}
\author{M. A. Kumar}
\affiliation{Department of Physics, University of California, Santa Barbara, CA 93106}
\author{Y. Zhang} 
\affiliation{Department of Physics, University of California, Santa Barbara, CA 93106}
\author{F. Yang}
\affiliation{Department of Physics, University of California, Santa Barbara, CA 93106}
\author{C. L. Tschirhart}
\affiliation{Department of Physics, University of California, Santa Barbara, CA 93106}
\author{M. Serlin}
\affiliation{Department of Physics, University of California, Santa Barbara, CA 93106}
\author{K. Watanabe} 
\affiliation{Research Center for Functional Materials,
National Institute for Materials Science, 1-1 Namiki, Tsukuba 305-0044, Japan}
\author{T. Taniguchi} 
\affiliation{International Center for Materials Nanoarchitectonics,
National Institute for Materials Science, 1-1 Namiki, Tsukuba 305-0044, Japan}
\author{A. H. MacDonald} 
\affiliation{Department of Physics, University of Texas, Austin, TX 78712}
\author{A. F. Young}   
\email{andrea@physics.ucsb.edu}
\affiliation{Department of Physics, University of California, Santa Barbara, CA 93106}

\maketitle
\textbf{Magnetism typically arises from the joint effect of Fermi statistics and repulsive Coulomb interactions, which favors ground states with non-zero electron spin. 
As a result, controlling spin magnetism with electric fields---a longstanding technological goal in spintronics and multiferroics\cite{matsukura_control_2015,jiang_electric-field_2018}---can be achieved only indirectly. 
Here, we experimentally demonstrate direct electric field control of magnetic states in 
an orbital Chern insulator\cite{sharpe_emergent_2019,serlin_intrinsic_2020,chen_tunable_2020, lu_superconductors_2019}, a magnetic system in which non-trivial band topology favors long range order of orbital angular momentum but the spins are thought to remain disordered\cite{xie_nature_2020,bultinck_anomalous_2019, zhang_twisted_2019,liu_correlated_2020,wu_collective_2020,chatterjee_symmetry_2019,repellin_ferromagnetism_2019,alavirad_ferromagnetism_2019}.  
We use van der Waals heterostructures consisting of a graphene monolayer rotationally faulted with respect to a Bernal-stacked bilayer to realize narrow and topologically nontrivial valley-projected moir\'e minibands\cite{ma_topological_2019, park_gate_2020, rademaker_topological_2020}.
At fillings of one and three electrons per moir\'e unit cell within these bands, we observe quantized anomalous Hall effects\cite{chang_experimental_2013} with transverse resistance approximately equal to $h/2e^2$ (where $h$ is Planck’s constant and $e$ is the charge on the electron), which  is indicative of spontaneous polarization of the system into a single-valley-projected band with a Chern number equal to two. 
At a filling of three electrons per moir\'e unit cell, we find that the sign of the quantum anomalous Hall effect can be reversed via field-effect control of the chemical potential; moreover, this transition is hysteretic, which we use to demonstrate nonvolatile electric field induced reversal of the magnetic state.
A theoretical analysis\cite{zhu_curious_2020} indicates that the effect arises from the topological edge states, which drive a change in sign of the magnetization and thus a reversal in the favored magnetic state. 
Voltage control of magnetic states can be used to electrically pattern nonvolatile magnetic domain structures hosting chiral edge states, with applications ranging from reconfigurable microwave circuit elements to ultralow power magnetic memory.}

The quantized anomalous Hall effect\cite{haldane_model_1988} occurs in two dimensional insulators whose filled bands have a finite net Chern number, and requires broken time-reversal symmetry.
Chern bands arise naturally in graphene systems when the Dirac spectrum acquires a mass, for instance due to the breaking of sublattice symmetry in monolayer graphene by a hexagonal boron nitride substrate\cite{song_topological_2015}.  
Absent electron-electron interactions, bands located in the two inequivalent valleys at opposite corners of graphene's hexagonal Brillouin zone are constrained by time reversal symmetry to acquire equal and opposite Chern numbers. 
In some graphene systems, a periodic moir\'e superlattice can be used to engineer superlattice bands which generically preserve the nonzero Chern numbers that arise from the incipient Berry curvature of the monolayer graphene Dirac points\cite{zhang_nearly_2019, liu_quantum_2019-1}. 
When the bandwidth of the superlattice bands is sufficiently small, the importance of electron-electron interactions is enhanced leading to symmetry breaking that manifests primarily as resistivity peaks at integer filling of normally four-fold degenerate superlattice bands\cite{bistritzer_moire_2011-1,cao_correlated_2018, chen_evidence_2019}. 
Among the states potentially favored by interactions are those with spontaneous breaking of time reversal symmetry\cite{bultinck_anomalous_2019, zhang_twisted_2019,liu_correlated_2020,wu_collective_2020,chatterjee_symmetry_2019,repellin_ferromagnetism_2019,alavirad_ferromagnetism_2019}, and indeed ferromagnetism and quantum anomalous Hall effects have been observed in both twisted bilayer graphene aligned to hexagonal boron nitride\cite{sharpe_emergent_2019, serlin_intrinsic_2020} and rhombohedral trilayer graphene, also aligned to hexagonal boron nitride\cite{chen_tunable_2020}. In contrast to quantum anomalous Hall effects observed in magnetically doped topological insulators which are spin ferromagnets rendered strongly anisotropic by their large spin orbit coupling, in graphene moir\'e systems the spin orbit coupling vanishes and the magnetism is thought to be primarily orbital, leading these systems to be dubbed `orbital Chern insulators'.

\begin{figure*}[ht!]
\includegraphics[]{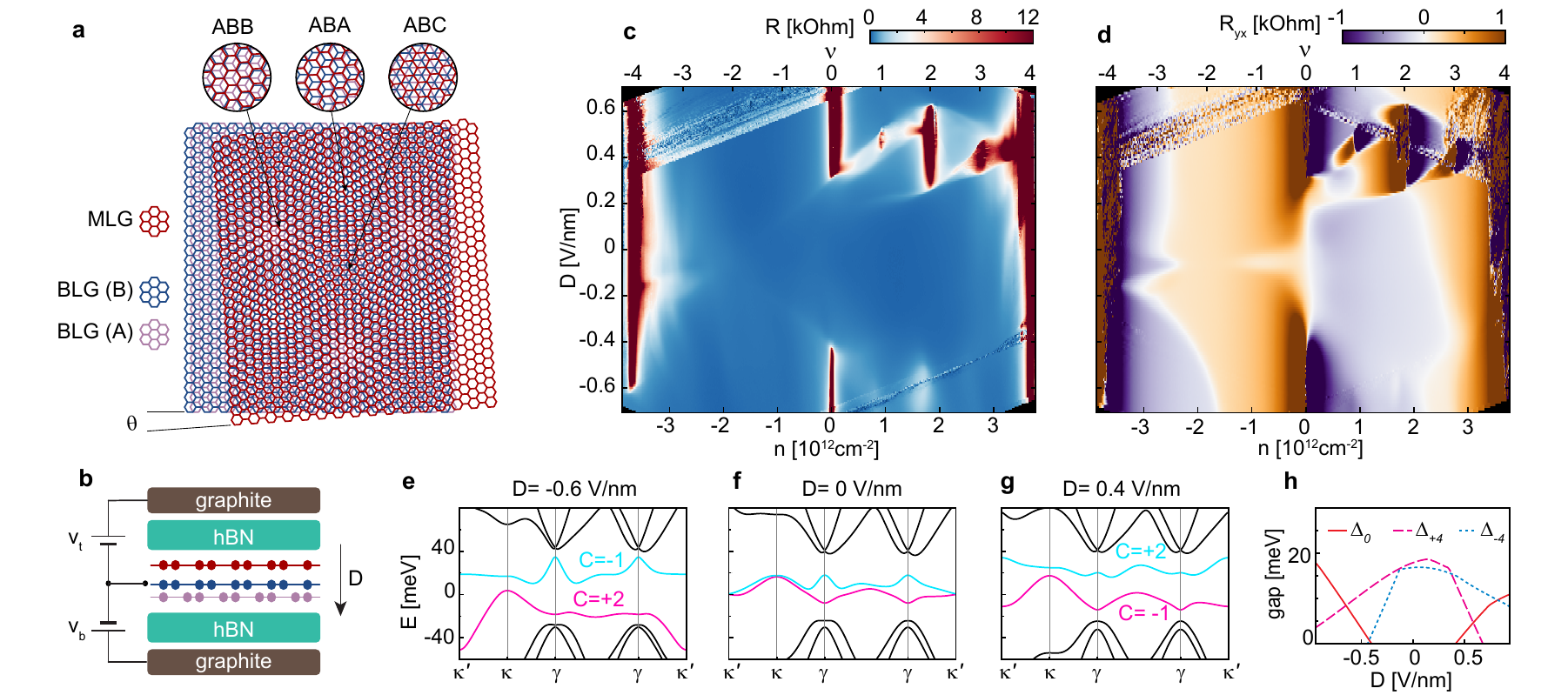} 
\caption{\textbf{Twisted monolayer-bilayer graphene.}
\textbf{a,} Crystalline structure of small angle twisted mono-bi twisted graphene. ABB regions form a triangular lattice separated by regions of ABA and ABC stacking. 
\textbf{b,} 
Schematic of our dual-gated devices. Top and bottom gate-voltages $\mathrm{v}_t$ and $ \mathrm{v}_b$  are used to control both total carrier density $n$ and  the  electric displacement field $D$ as described in the main text. \textbf{c,} Longitudinal resistance $R_{xx}$ measured at T=1.35~K and \textbf{d,} transverse resistance $R_{yx}$ measured at T=1.35~K and $B=\pm0.5$~T. Both are plotted as a function of  carrier density $n$ and $D$ for device D1 with twist angle $\theta= 1.25^\circ$.
\textbf{e,} Band structure calculated from a continuum model (see Methods and SI) for displacement field $D=-0.6$~V/nm.
\textbf{f,} Band structure for $D=0$
 and \textbf{g,} D=0.4~V/nm.
\textbf{h,} Energy gaps calculated within the Hartree approximation (see SI) for $\nu=- 4, 0$, and $+4$. 
In \textbf{e-h}, we use a dielectric constant $\epsilon_{\text{bg}}=4$ to convert the interlayer potential difference to a displacement field, $\Delta_U = e D d/\epsilon_{\text{bg}}$, where d=3.3~\AA\ is the graphene interlayer separation.}

\label{fig:1}
\end{figure*}

\begin{figure*}[ht!]
\includegraphics[]{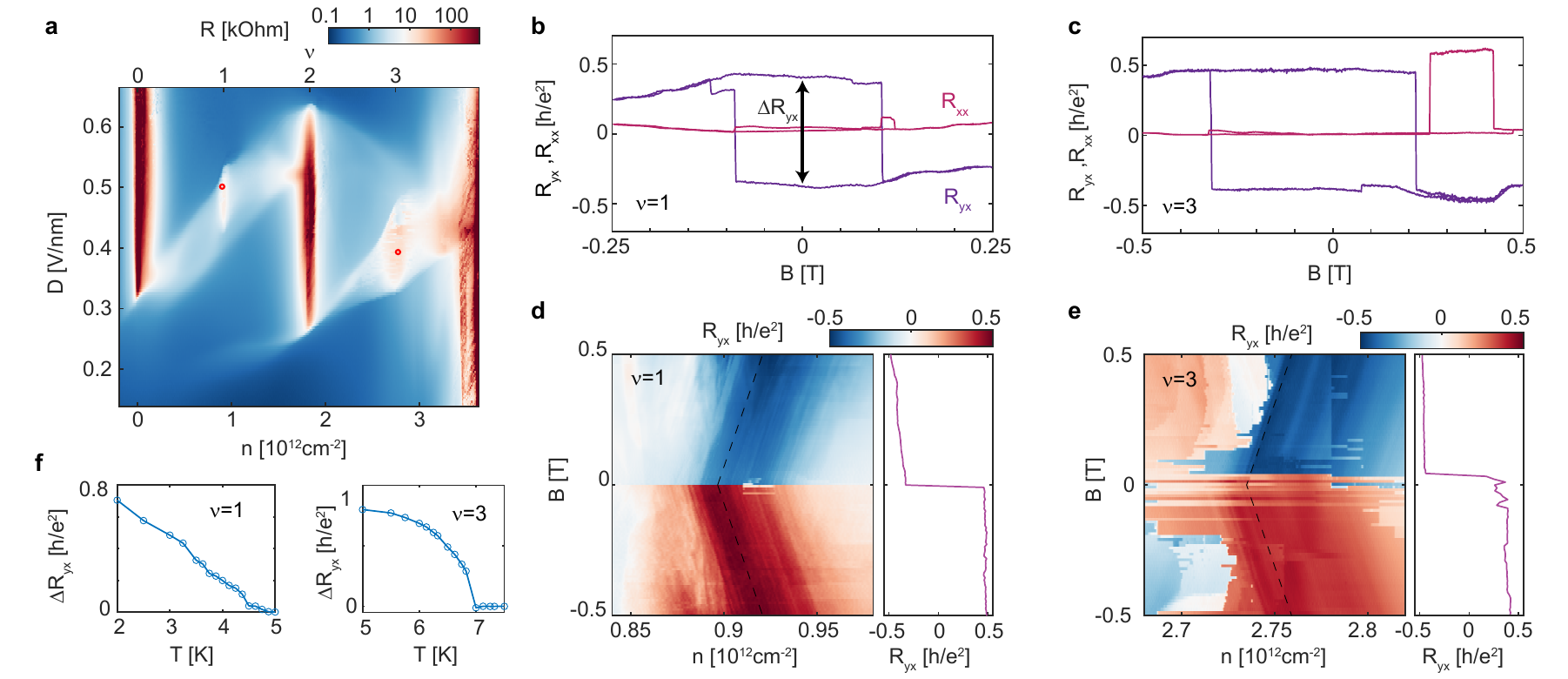} 
\caption{\textbf{Orbital Chern insulators with C=2.}
\textbf{a,} High resolution plot of $R_{xx}$ measured at T=1.35~K and B=0 in device D1. 
\textbf{b,} $R_{yx}$ and  $R_{xx}$ measured as a function of $B$ at 1.43~K near $\nu=1$, at n=0.9$\times 10^{12}\mathrm{cm}^{-2}$ and $D=0.5$~V/nm, and \textbf{c,} near $\nu=$3, at n=2.77$\times 10^{12}\mathrm{cm}^{-2}$  and $D=0.39$~V/nm.
\textbf{d,} $B$ and $n$ dependence of $R_{yx}$ near $\nu=1$ measured at T=20~mK and 
(\textbf{e}) near $\nu=3$ measured at T=1.35~K. In these measurements the fast sweep axis is $n$. Tilted dashed lines show the slope expected for gaps associated with Chern number $C=2$, matching the $n-B$ evolution of the plateaus in $R_{yx}$. 
Insets at right in panels d and e shows the $R_{yx}$ plotted along the dashed lines.
\textbf{f,} Temperature dependence of the  hysteresis loop height at $B=0$, $\Delta R_{yx}$ for $\nu=1$ and $\nu=3$.  Hysteresis vanishes at $T_C\approx 5$~K and $T_C\approx 7$~K, respectively, defining a lower bound for the Curie temperature.
}
\label{fig:2}
\end{figure*}

Here we introduce another moir\'e heterostructure that shows quantum anomalous Hall effects.  As shown in Figure 1a, our devices consist of a graphene monolayer rotationally faulted with respect to a Bernal stacked graphene bilayer, which we refer to as twisted monolayer-bilayer graphene (tMBG). The tMBG moir\'e consists of a triangular lattice of ABB-stacked regions interspersed with more structurally stable ABA and ABC regions, as illustrated 
in Figure~\ref{fig:1}a; in the low energy bands, wave functions are localized near the ABB regions. Our devices are fabricated by applying a `cut-and-stack' method to an exfoliated graphene flake that contains both monolayer and bilayer graphene regions (see Methods and Extended Data Fig.~\ref{fig:S:fab}). Two graphite gates above and below the tMBG layer allow independent control of the overall carrier density $n=c_t\mathrm{v}_t+c_b \mathrm{v}_b$ and electric displacement field, $D=\left(c_t\mathrm{v}_t-c_b\mathrm{v}_b\right)/2\varepsilon_0$, where $\varepsilon_0$ is the  vacuum permittivity, $\mathrm{v}_{t(b)}$ is the applied voltage and $c_{t(b)}$ is the capacitance per unit area of the top (bottom) gate (see Figure~\ref{fig:1}b).  

\subsection*{D-field tunable flat bands}
Figs.~\ref{fig:1}c-d show longitudinal and Hall resistance as a function of $n$ and $D$ for a device with interlayer twist angle $\theta\approx 1.25^\circ$. 
Additional data from devices with a range of twist angles between 0.9 and 1.4$^\circ$ are shown in Extended Data Fig.~\ref{fig:S:alldevices}. 
All devices show resistance peaks at $\nu=\pm 4$, where $\nu=n \auc$ denotes the number of carriers per superlattice unit cell,  $\auc\approx\sqrt{3}a^2/(2 \sin^2\theta)$ is the unit cell area and $a=2.46$~\AA\ is the lattice constant of graphene.  
Correlated states, revealed by sign changes in the Hall resistance and peaks in the longitudinal resistance are observed near  $\nu=1,2$ and 
$3$ for devices with $\theta=1.25^\circ$ and $1.4^\circ$.  They appear only within a narrow range of displacement fields near $D\approx 0.4$~V/nm, with onset temperatures as high as 20~K (Figure \ref{fig:S:tdepR}).  In all such devices an anomalous Hall effect is observed at $\nu=1$ and $\nu=3$, as shown in \ref{fig:S:hystMK6} and \ref{fig:S:hystGP45}.  

Similar observations in rotationally faulted bilayer-bilayer graphene\cite{shen_correlated_2020, liu_spin-polarized_2019, cao_electric_2019, burg_correlated_2019, he_tunable_2020}
have been interpreted as arising from displacement field tuned formation of an isolated narrow band,
and consequent spontaneous breaking of spin, valley, or lattice symmetries that result in correlated insulating states at integer fillings.  
A previous theoretical study\cite{ma_topological_2019} has  suggested that tMBG similarly hosts narrow electronic bands at small twist angles around $1.2^\circ$. 
In our data, the domains of displacement field over which
correlated physics is observed aligns 
with where numerical simulations (Figure~\ref{fig:1}e-h) 
show the formation of a narrow, isolated band.  
The electronic band structure of tMBG arises from the moir\'e-induced hybridization of the monolayer graphene Dirac cone 
(located at the $\kappa$ point in the band structure diagrams of Figs. 1e-g) with the parabolic low energy band of the 
bilayer (at the $\kappa '$ point).  
When the displacement field $D=0$, low-energy valence and conduction bands are isolated from the other bands but overlap with each other, giving rise to gaps at $\nu=\pm 4$ but not at $\nu=0$. 
At intermediate values of $|D|$ a gap opens at charge neutrality leading to the formation of isolated conduction and valence bands with nonzero Chern number. Further increase of $|D|$ then leads to band overlap between the low energy bands and higher energy dispersive bands.  
Correlated states at integer filling, accompanied by `halos' of changed resistance relative to the background at non-integer filling, 
appear in the intermediate regime where both calculations (Figure 1h) and experimental data indicate the formation of a narrow and isolated conduction band.

\subsection*{Quantum anomalous Hall states with C=2}
We focus on the $n>0$, $D>0$ narrow-band regime, a detail of which is shown in Figure 2a.  At $\nu=2$, we observe a robust insulator, consistent with a topologically trivial gap. At both $\nu=1$ and $3$, however, the resistance at B=0 is both noisy and comparatively low.  
Magnetoresistance measurements (Figs. 2b-c) reveal that the noise is due to magnetic hysteresis.  
In both cases, we observe rapid switching between states with $R_{yx}\approx \pm h/2e^2$, accompanied by a low $R_{xx}\lesssim 1$~$\mathrm{k\Omega}$.   The saturation of the Hall resistance near $h/2e^2$ is suggestive of polarization into bands with Chern number 2.  
The quantization is not precise, reaching only 85\% of the expected value at $\nu=3$, for instance. However, the evolution of the Hall plateau in magnetic field, shown in Figs.~2d-e, provides further evidence of an underlying orbital Chern insulator state. For both $\nu=1$ and 3, increasing the magnetic field shifts the center of the plateau in density ($n^*$), in agreement with the St\v{r}eda formula\cite{streda_quantised_1982}, $C=(h/e)\partial n^*/\partial B$, which is applicable to any incompressible Chern insulator.
Chern 2 bands were previously predicted\cite{ma_topological_2019} and are consistent with our own band structure calculation, which indicated that the conduction band has C=2 for a positive $D$ field.  

It is notable that tMBG, in contrast to both rhombohedral trilayer graphene and twisted bilayer graphene, does not rely on precise alignment to hexagonal boron nitride making it an all-carbon quantum anomalous Hall system.  However, our observations at $\nu=1$ and $\nu=3$ are in many ways qualitatively similar to the 
quantum anomalous Hall effect characterizations reported previously\cite{sharpe_emergent_2019, chen_tunable_2020, serlin_intrinsic_2020}.  
For instance, the magnetic transitions tend to occur via several discrete steps, corresponding to a small number of micron-sized mesoscopic domain reversals.
Temperature dependent measurements show Curie temperatures---defined here by the onset of hysteresis---of $T_C\approx 5$~K for $\nu=1$ and $T_C\approx 7$~K for $\nu=3$ (see Figures \ref{fig:2}f  as well as \ref{fig:S:tdephyst}-\ref{fig:S:tdepah}), again similar to previous reports of ferromagnetism in moir\'e heterostructures\cite{sharpe_emergent_2019,chen_tunable_2020, serlin_intrinsic_2020}. 
We expect the lack of perfect quantization to arise from disorder-induced domain structure, as recently observed for similar states in twisted bilayer graphene\cite{tschirhart_imaging_2020}.
\begin{figure*}[ht!]
\includegraphics[]{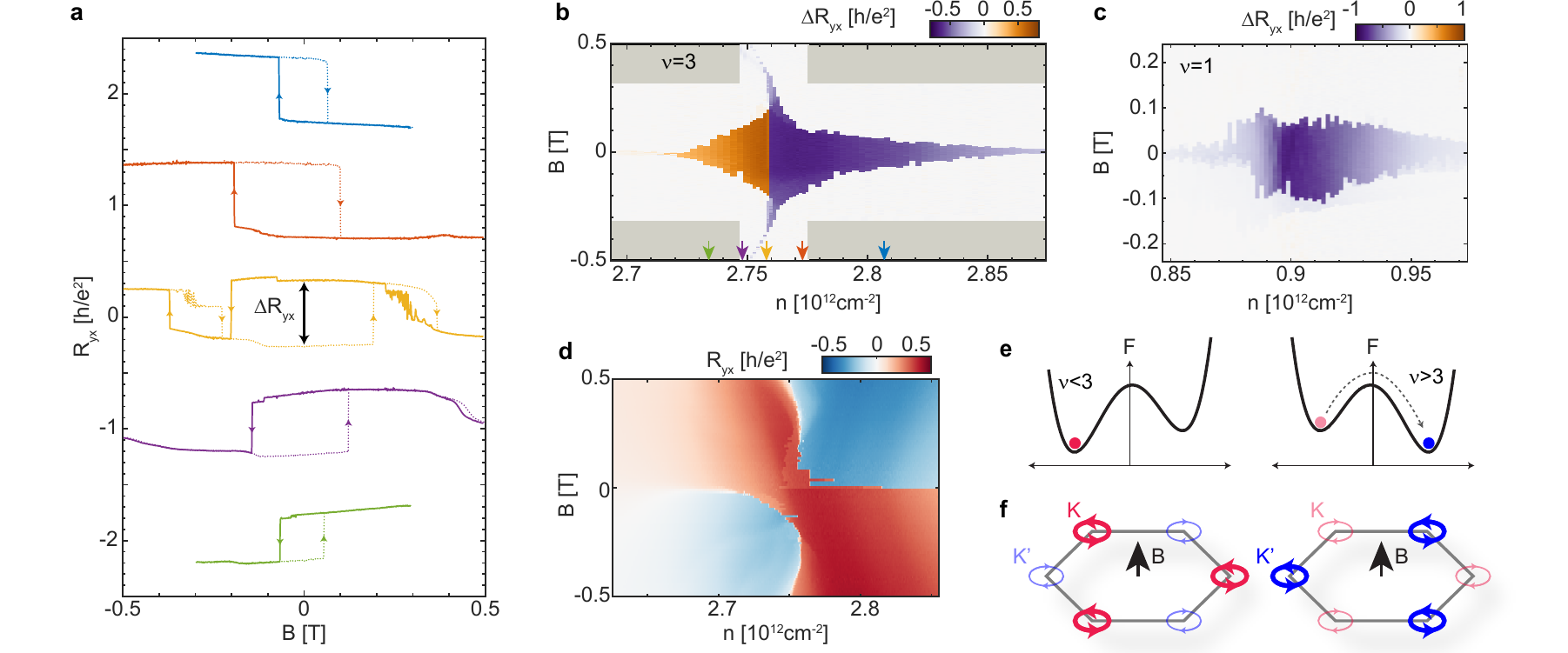} 
\caption{\textbf{Doping induced magnetization reversal.}
\textbf{a,} Hysteresis loops measured at $T$=6.4~K for several densities near $\nu=3$. From top to bottom, n=2.807, 2.773, 2.758, 2.748, 2.734 $\times 10^{12}$cm$^{-2}$, with colors corresponding to arrows in panel (b). Curves are offset by $h/e^2$.
\textbf{b,} $\Delta R_{yx}=R_{yx}^{B_{\downarrow}} - R_{yx}^{B_{\uparrow}}$ near $\nu$=3 measured at 6.4K.
\textbf{c,} $\Delta R_{yx}$  near $\nu$=1 measured at $T\approx 20$~mK.
\textbf{d,} $R_{yx}$ as a function of field and doping at 6.4~K with doping the fast sweep direction.
\textbf{e,} Schematic diagram of the free energy $F$ {\it versus} order parameter for 
the two magnetic states. For a fixed sense of valley polarization, the orbital magnetization 
reverses sign with doping. \textbf{f,} At fixed magnetic field the systems therefore 
switches between $K$ and $K'$ valley polarization. 
}
\label{fig:3}
\end{figure*}

\subsection*{Magnetization reversal}
Striking new phenomena are, however, observed in the $n$-dependence of the Hall effect.   
Specifically, whereas $R_{yx}$ changes smoothly as a function of $n$ near the $\nu=1$ orbital Chern insulator, 
it exhibits erratic switching behavior near $\nu=3$ (Figure~2d,e). 
To investigate this phenomenon, we perform a dense series of hysteresis loop measurements in the vicinity of  $\nu=3$ at higher temperature, where we expect that domain wall pinning is weaker and 
hysteretic effects are somewhat suppressed. While the sign of the anomalous Hall resistance is constant near $\nu=1$, it reverses abruptly upon crossing $\nu=3$(see Figs.~\ref{fig:3}a-c).
The sign reversal occurs with minimal change in the magnitude of $R_{yx}$, which remains close to the quantized value.  
This suggests that the reversal occurs via a change in the product of the magnetization
sign and the Chern number sign, which we refer to as the magnetic state.  
This is further evidenced by the $n$-$B$ map of $R_{yx}$ 
(see Figs.~\ref{fig:3}d), which shows that the $R_{yx}$ changes sign at 
$n\approx2.76\times 10^{12}\mathrm{\mathrm{cm}^{-2}}$, corresponding to $\nu=3$. 
An additional manifestation of the inversion of the sign of the magnetization of
a given magnetic state is the abrupt upturn in the coercive field in the close vicinity of the 
reversal point shown in Figure~\ref{fig:3}b.  
This phenomenon is qualitatively consistent with a picture in which the total magnetization $M$
changes sign for a fixed sense of valley polarization by passing through zero; near $M=0$, the coupling to the magnetic field vanishes leading to a divergence of the coercive field. 

\begin{figure*}[ht!]
\includegraphics[]{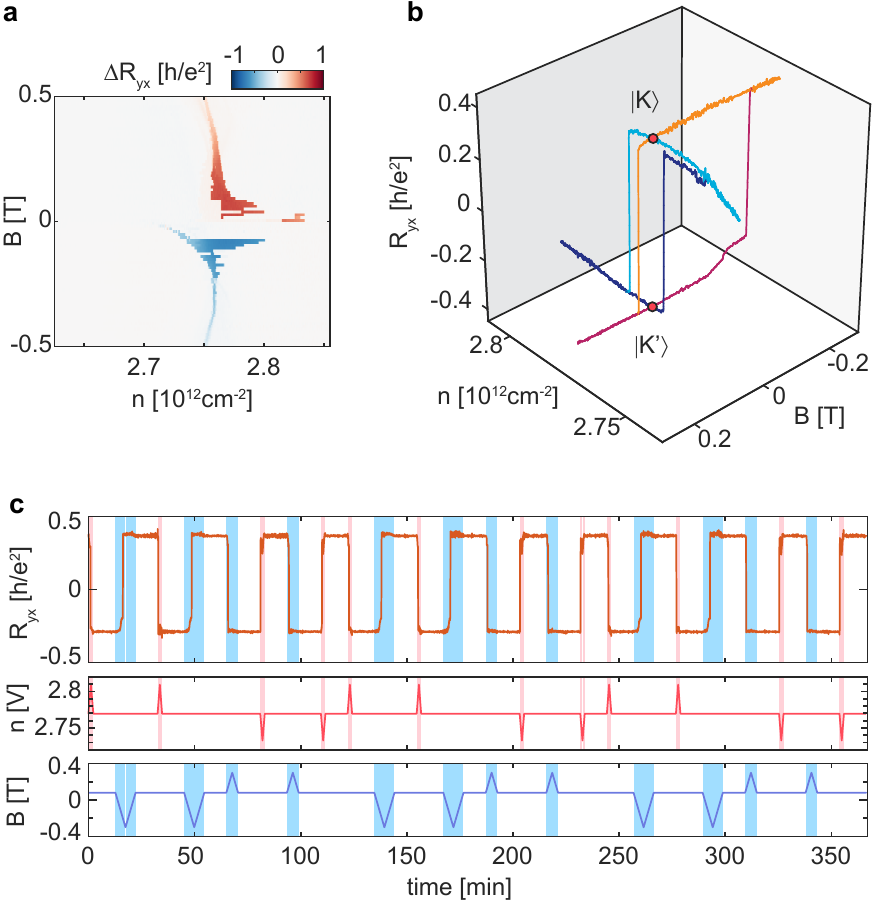} 
\caption{\textbf{Nonvolatile electrical control of a magnetic state at T=6.4K.}
\textbf{a,} Electric field hysteresis.  Color scale encodes the difference between trace and retrace as the density is swept using electrostatic gates, with each trace-retrace taken at fixed magnetic field.  Reproducibility of magnetization switches is shown in Figure \ref{fig:S:hystrepeat}.  
\textbf{b,} Examples of hysteresis loops in both $n$ and $B$.  
\textbf{c,} Time series of alternating pulses of field-effect density and magnetic field showing reproducible and nonvolatile switching of the magnetic state, as read by the resulting Hall resistance.
}
\label{fig:4}
\end{figure*}

We propose that the magnetization reversal arises from the unique features of orbital Chern insulators.  In particular, the protected edge states required by their nontrivial topology are occupied and contribute to the magnetization when the insulator is slightly $n$-doped, but not when it is 
slightly $p$-doped\cite{zhu_curious_2020}.  The edge state contribution 
leads to a jump $\delta M$ in the magnetization across the Chern insulator gap, 
\begin{equation}
\delta M =C\frac{\Delta}{\pi \hbar^2/m_e \auc} \cdot \frac{\mu_B}{\auc}\label{deltam}
\end{equation}
where $m_e$ is the electron mass, $\auc$ is the unit cell area, $\Delta$ is the gap, and $\mu_B$ is the Bohr magneton. 
While this effect is present in all Chern insulators, it is negligible in magnetically doped topological insulators\cite{chang_experimental_2013} where time-reversal symmetry is broken primarily by spontaneous spin-polarization.  
In these systems, the total spin magnetization is of order $1 \mu_B/\auc$.
Since $\Delta$ is on the order of a few meV, $\Delta\ll\hbar^2/(m_e \auc)$ and  the edge-state orbital magnetization is dwarfed by spin magnetization.   
In graphene moir\'e orbital Chern insulators the bulk orbital magnetization is again of order $1 \mu_B/\auc$ \cite{zhu_curious_2020,tschirhart_imaging_2020}.
However, the large  unit cell area allows for the prefactor in Eq. \ref{deltam} to be of comparable or larger magnitude.
When this edge state contribution is sufficiently large, it can lead to a reversal in the \textit{sign} of the net magnetization across the energy gap.  For instance, taking measured energy gaps  for quantum anomalous Hall states in twisted bilayer graphene\cite{serlin_intrinsic_2020} produces an estimate of $\delta M\approx \mu_B/\auc$.
Indeed, a reexamination of transport data from the tBLG device aligned to hexagonal boron 
nitride studied in Ref.~\onlinecite{serlin_intrinsic_2020} finds a similar, though much less dramatic, change in the sense of the hysteresis loop accompanied by a similar divergence of the coercive field near (though not precisely at) $\nu=3$ (see Figure \ref{fig:s:tblg}).  We note that the magnetization of an orbital magnet can in principle change sign at any filling factor, but that it is more likely at integer $\nu$ because the magnetization is discontinuous at this density.The conditions under which  the reversal occur are discussed in Methods.

\subsection*{Electrical switching of magnetic states}
Because an external magnetic field favors the state with magnetization aligned with the field, changing the sign of $M$ is predicted to drive a reversal of the valley polarization as the chemical potential crosses the gap (see Figure \ref{fig:3}e\change{}{-f}).  
The quantized Hall effect sign reversal is thus due to a change in magnetic state, in contrast to previously observed anomalous Hall effect sign reversals\cite{chiba_anomalous_2010,zhang_experimental_2020} arising from a change in the sign of the total anomalous Hall effect for a \textit{given} 
magnetic state.  As a result, electric field induced hysteretic behavior is possible in our case since 
the two states are separated by an energy barrier. Fig. \ref{fig:4}a depicts the difference (again denoted $\Delta R_{yx}$) between traces and retraces of field-effect tuned density, acquired as $B$ is stepped.   Near $\nu=3$, $R_{yx}$ shows a finite signal in this channel, indicating hysteresis in gate voltage sweeps at fixed magnetic fields as high as several hundred millitesla.  

Electric field control of magnetic states provides a reliable experimental knob to realize nonvolatile switching of magnetization.  We find that at 80 mT, $n$=$2.77 \times 10^{12}$~cm$^{-2}$, and $T=6.4$~K, states of opposite polarization can be controllably switched using either excursions in $n$ \textit{or} $B$, as shown in Figure \ref{fig:4}b.  
Figure \ref{fig:4}c shows this principle applied to nonvolatile switching of the
magnetic state using both $B$ excursions at constant $n$ and $n$ excursions at constant $B$.  
Switches occur with perfect fidelity and appear to be indefinitely nonvolatile at this temperature.  We note that as for current- and magnetic field-driven switching, the excellent reproducibility of field-effect switching ultimately arises from the absence of states with partial or intermediate valley polarization, or equivalently the extreme anisotropy inherent in a purely orbital two dimensional magnet.  

Orbital magnets realized at room temperature would be immediately applicable as embedded magnetic memory in logic devices.  
Candidates for higher temperature operation include other moir\'e systems in which the lattice constant is smaller and the correlation energy consequently larger, assembled either by van der Waals stacking or using non-epitaxial growth techniques\cite{beekman_ferecrystals_2014}.
Even restricted to cryogenic temperatures, there exist several immediate opportunities that leverage gate switchable chirality in a quantum anomalous Hall system. As a simple example, orbital Chern insulators could be used as the central elements in reconfigurable and compact microwave circulators\cite{viola_hall_2014}, which may be useful for scaling up quantum information processing. 
Meanwhile, integrating orbital Chern insulators with superconductors may permit new device architectures 
devoted to the detection and manipulation of extended Majorana zero modes\cite{lian_topological_2018}.  
Our results highlight the novel opportunities for controlling functionality that arise from the realization of purely orbital magnetic systems. 

\section*{Methods}
\subsection*{Device fabrication}
Van-der-Waals heterostructures for tMBG devices are fabricated by a dry transfer technique based on a polycarbonate film on top of a polydimethyl siloxane (PDMS) stamp.  
The heterostructure is assembled in two steps to minimize the  chance to disturb the twist angle at the critical monolayer-bilayer interface. First a two layer stack  with hBN at the top and graphite layer at the bottom is picked up and transferred to a bare $\mathrm{Si/SiO_2}$ wafer. This partial stack is annealed in vacuum at $400^{\circ}$~C to remove polymer residues from the top surface. 
A second stack consisting (from top to bottom) of hBN-FLG-hBN-MLG-BLG(where FLG, MLG and BLG are few-layer, monolayer and bilayer graphene, respectively) is then assembled and transferred on top of the first  one. Crucially, the angle registry between the MLG and BLG is ensured by starting from a single exfoliated flake that includes both MLG and BLG domains.  The flake is then cut with a conducting atomic force microscope tip in air (see Extended Data Fig.~\ref{fig:S:fab}). The MLG piece is  picked up using a PDMS stamp and an interlayer twist introduced by substrate rotation, and then the  BLG pieced picked up. 
All flakes except MLG and BLG are picked up at approximately $90^{\circ}$~C; the MLG, BLG, and the transfer to the final substrate are made at $30^{\circ}$~C to preserve the twist angle of the structure. Devices are then fabricated using standard e-beam lithography and $\mathrm{CHF_3/O_2}$ etching, and edge contacts are made by deposition of Cr/Pd/Au (1.5/15/250~nm). 

\subsection*{Device characterization}
Transport measurements were performed using SRS lock-in amplifier, scurrent preamplifiers (DL Instruments) and voltage preamplifiers (SRS) with typical excitations in the range 1-10~nA at 17.77~Hz.  Measurements were performed in either a cryogen free dilution refrigerator, with the sample in vacuum, or in the case of Extended Data Fig.~\ref{fig:S:tdepR}, in a wet variable temperature insert with the sample in helium vapor. 

For devices D1 and D2, the twist angle was  determined from the position of the correlated insulating states. For devices D3 and D4 the twist angle was determined from the periodicity of the Hofstadter features observed in magnetic field that arise from the interplay of the moir\'e superlattice period and magnetic length.

\subsection*{Band structure simulations}
Band structure simulations are performed within a continuum model analogous to that of Reference\cite{bistritzer_moire_2011-1} for the coupling between the top and middle layers while the middle and bottom layers are treated as Bernal-stacked bilayer graphene. More details are in the Supplementary Information (SI).

\subsection*{Electrical reversibility}
Why is the field-effect reversibility so robust in tMBG at $\nu=3$, but weak in twisted bilayer graphene at $\nu=3$ and absent at other filling factors in either system? 
Field reversibility requires that the edge state contribution to the magnetization, $\delta M$, reverses the sign of the total magnetization. Our calculations for both tMBG and twisted bilayer graphene aligned to hBN indicate that $\delta M$ and $M_{bulk}$--defined as the total magnetization when the chemical potential is at the bottom of the gap---normally have opposite 
sign. To ensure field reversibility, one thus wants to maximize the magnitude of the edge state contribution while minimizing the magnitude of the bulk contribution. 
The large Chern numbers in tMBG are certainly helpful in this regard, as they increase $\delta M$.  
In addition, our theoretical analysis indicates that in odd layer systems, exchange splitting between bands arising from valley polarization further decreases the magnitude of $M_{bulk}$.  This effect, which includes contributions from bands far from the Fermi level, is present in tMBG but absent in twisted bilayer (see SI for details). 
The exchange contribution is expected to be larger at $\nu=3$ than at $\nu=1$, possibly explaining why we see
reversible Chern insulators at $\nu=3$ but conventional Chern insulators at $\nu=1$ in tMBG.

%

\section*{acknowledgments}
The authors acknowledge discussions with  J. Checkelsky, S. Chen, C. Dean, M. Yankowitz, D. Reilly, I. Sodemann, and M. Zaletel for discussions. Work at UCSB was primarily supported by the ARO under MURI W911NF-16-1-0361.  
Measurements of twisted bilayer graphene (Extended Data Fig. \ref{fig:s:tblg}) and measurements at elevated temperatures (Extended Data Fig.~\ref{fig:S:tdepR}) were supported by a SEED grant and made use of shared facilities of the UCSB MRSEC (NSF DMR 1720256), a member of the Materials Research Facilities
Network (www.mrfn.org).  
AFY acknowledges the support of the David and Lucille Packard Foundation under award 2016-65145.
AHM and JZ were supported by the National Science Foundation through the Center for Dynamics and Control of Materials,  an NSF MRSEC under Cooperative Agreement No. DMR-1720595, and by the Welch Foundation under grant TBF1473.
CLT acknowledges support from the Hertz Foundation and from the National Science Foundation Graduate Research Fellowship Program under grant 1650114.  
KW and TT acknowledge support from the Elemental Strategy Initiative
conducted by the MEXT, Japan, Grant Number JPMXP0112101001,  JSPS
KAKENHI Grant Numbers JP20H00354 and the CREST(JPMJCR15F3), JST.

\section*{Author contributions}
HP, MAK, and YZ fabricated the devices.  HP, FY, CLT and MS performed the measurements, advised by AFY.  JZ and AHM performed the band structure calculations.  KW and TT grew the hexagonal boron nitride crystals.  HP, AHM, and AFY wrote the manuscript with input from all other authors. 

\section*{Competing interests}
The authors declare no competing interests.

\section*{Data availability}
Source data are available for this paper. All other data that support the plots within this paper and other findings of this study are available from the corresponding author upon reasonable request.




\renewcommand{\figurename}{\textbf{Extended Data Fig.}}
\renewcommand{\thefigure}{E\arabic{figure}}
\setcounter{figure}{0}

\begin{figure*}[ht!]
\includegraphics[]{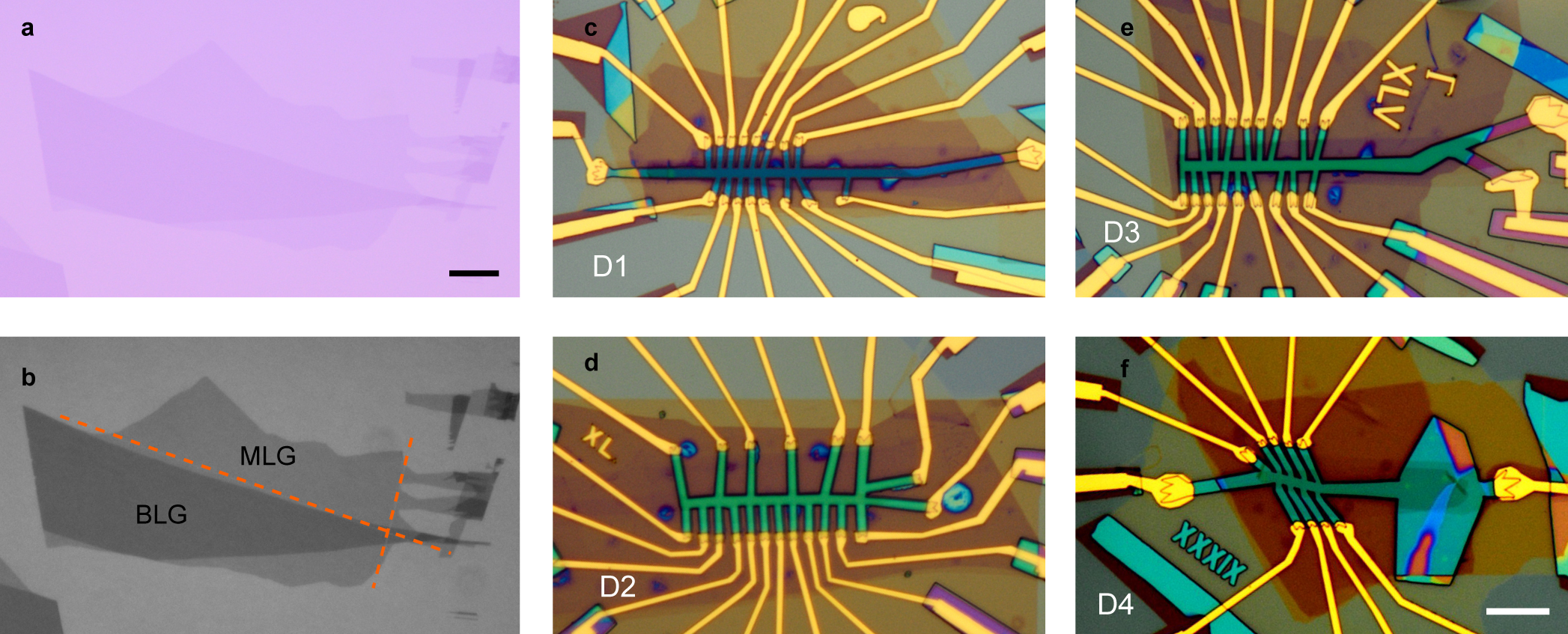} 
\caption{\textbf{tMBG devices.}
\textbf{a, b} optical image of a typical graphene flake containing both MLG and BLG domains.  Dashed lines in (b) show the lines along which the flake was cut using AFM.
\textbf{c, d, e, f} Optical images of completed tMBG devices D1, D2, D3, D4. Scalebar is 10 ~$\mathrm\mu m$.
}
\label{fig:S:fab}
\end{figure*}

\begin{figure*}[ht!]
\includegraphics[]{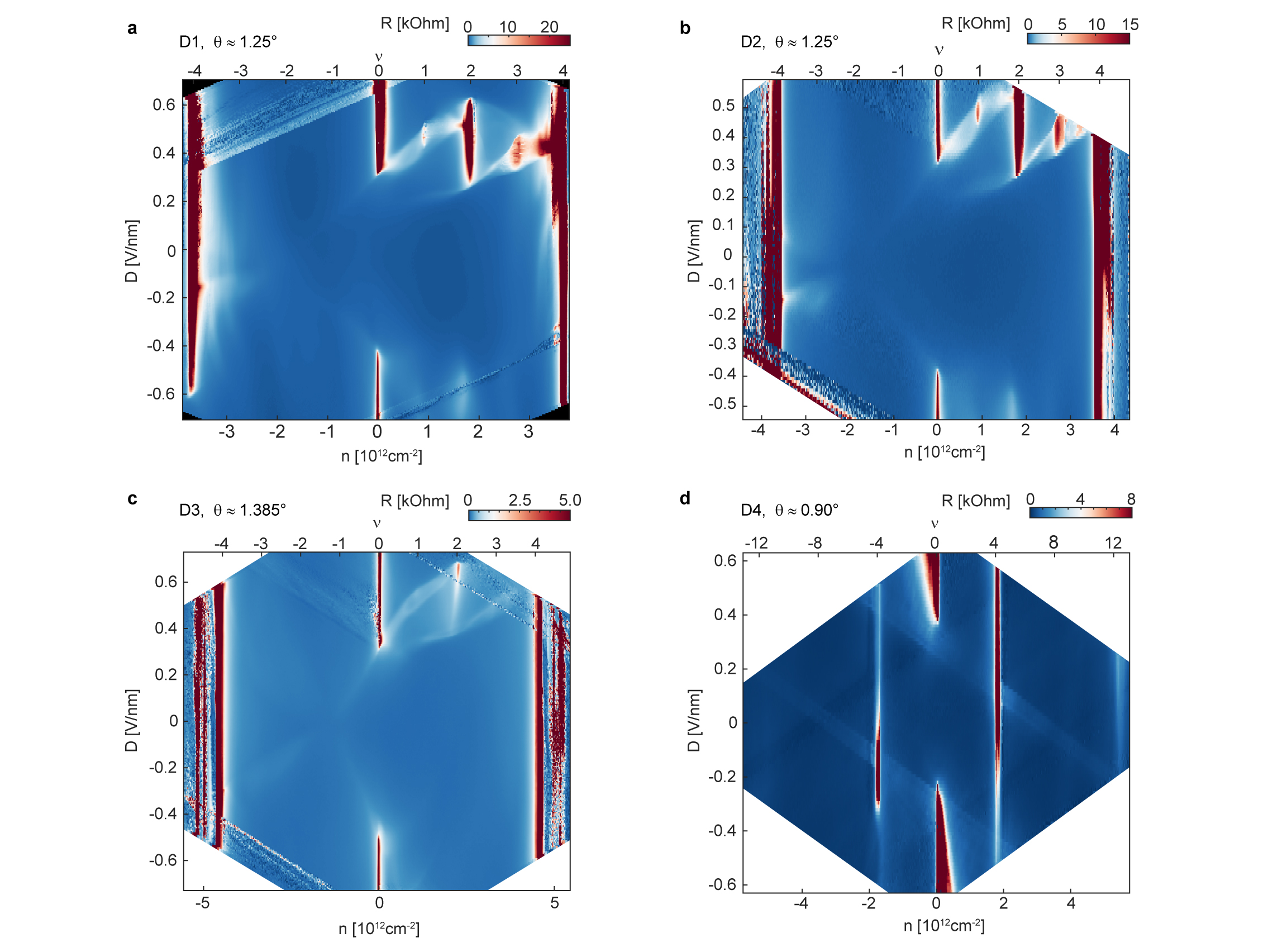} 
\caption{\textbf{Longitudinal resistance of tMBG devices with different twist angles.}
Longitudinal resistance $ R_{xx}$ of devices \textbf{a,} D1 with $\theta\approx1.25(1)^\circ$, \textbf{b,} D2 with $\theta\approx1.25(1)^\circ$, \textbf{c,} D3 with $\theta=1.385(5)^\circ$, and \textbf{d,} D4 $\theta\approx0.90(1)^\circ$. Here the number in parentheses indicates uncertainty in the final digit.  All measurements are performed at zero magnetic field and $T\approx 20$~mK.
}
\label{fig:S:alldevices}
\end{figure*}

\begin{figure*}[ht!]
\includegraphics[]{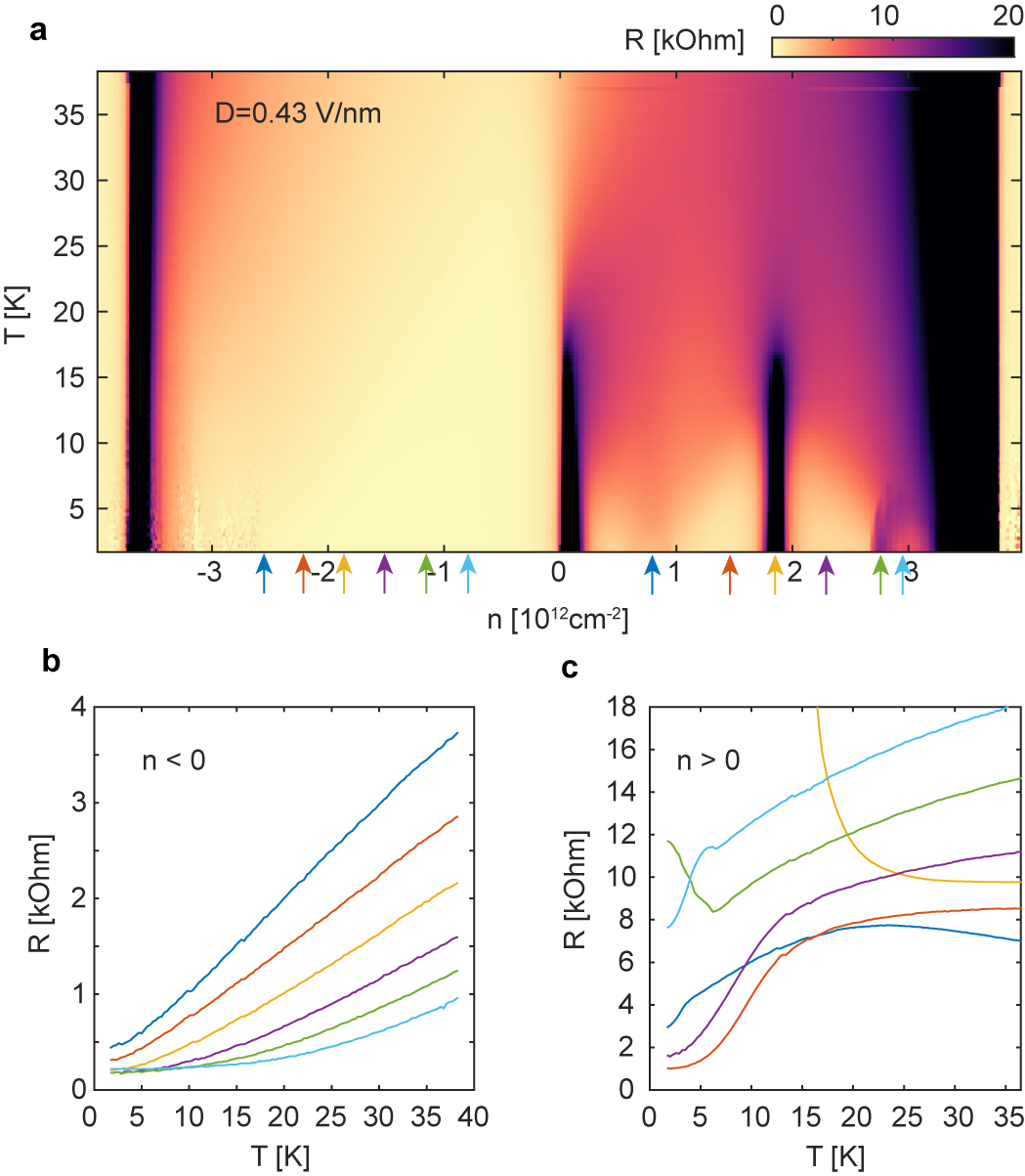} 
\caption{\textbf{Temperature dependence of the correlated states in device D1.}
\textbf{a}.  Temperature dependent resistance measured at D=0.43~V/nm in device D1.
\textbf{b, c}  Temperature-dependent resistance at selected carrier densities, marked by arrows in  (a).
}
\label{fig:S:tdepR}
\end{figure*}

\begin{figure*}[ht!]
\includegraphics[]{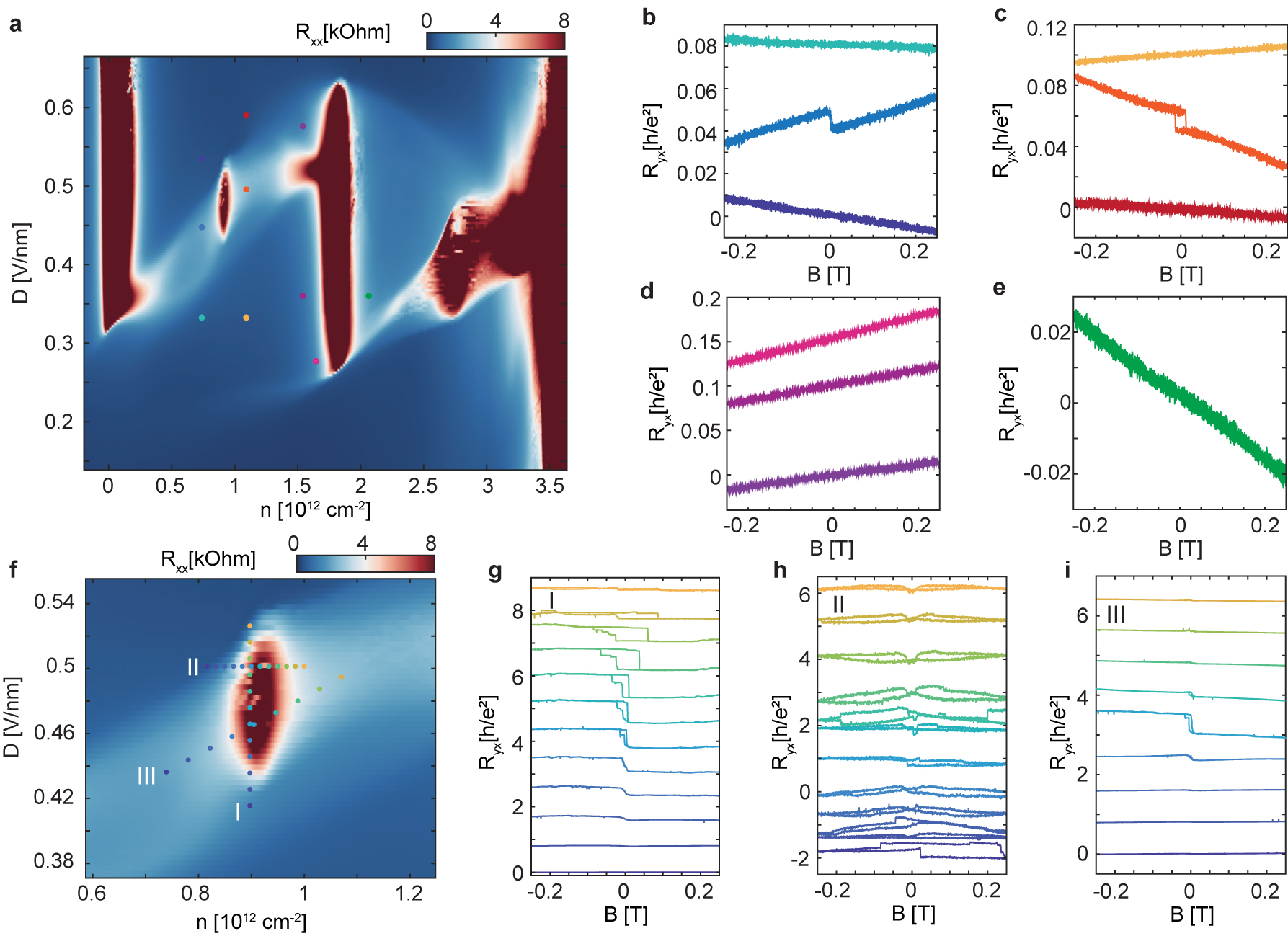} 
\caption{\textbf{Development of Hall resistance at different $n$ and $D$ in device D1.}
\textbf{a.} Longitudinal resistance $R_{xx}$ of the correlated region at $B=0$ T. \textbf{b-e.} Hall resistance $R_{yx}$ measured at $n$ and $D$ marked by dots in \textbf{a}. \textbf{f.} Zoom in of $R_{xx}$ around $\nu=1$. $R_{yx}$ measured along the line cut I, II, and III are present in \textbf{g-i}. $R_{yx}$ in the plots are shifted by an offset.
}
\label{fig:S:hystMK6}
\end{figure*}

\begin{figure*}[ht!]
\includegraphics[]{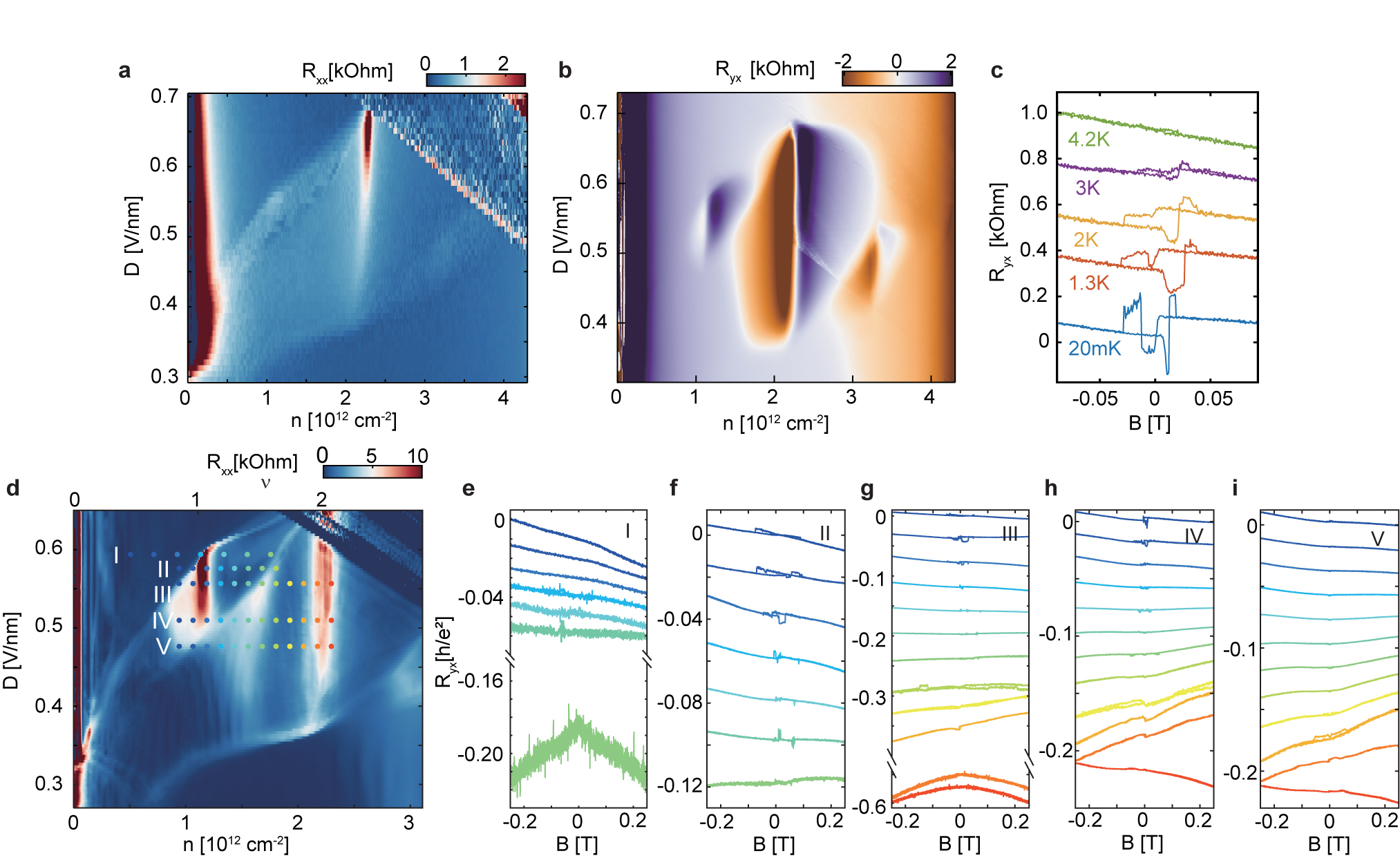} 
\caption{\textbf{Development of Hall resistance in device D3.}
\textbf{a.} Longitudinal resistance $R_{xx}$ of the correlated region at $B=0$ T. \textbf{b.} Hall resistance $R_{yx}$ of the same region as in \textbf{a}. \textbf{c.} Temperature dependence of $R_{yx}$ at $\nu=1$. The anomalous Hall effect disappears at $4.2$ K. \textbf{d.} $R_{xx}$ of the correlated region measured at $B=2$ T. $R_{yx}$ along the line cuts I-V are plotted in panels \textbf{e-i}.
}
\label{fig:S:hystGP45}
\end{figure*}

\begin{figure*}[ht!]
\includegraphics[]{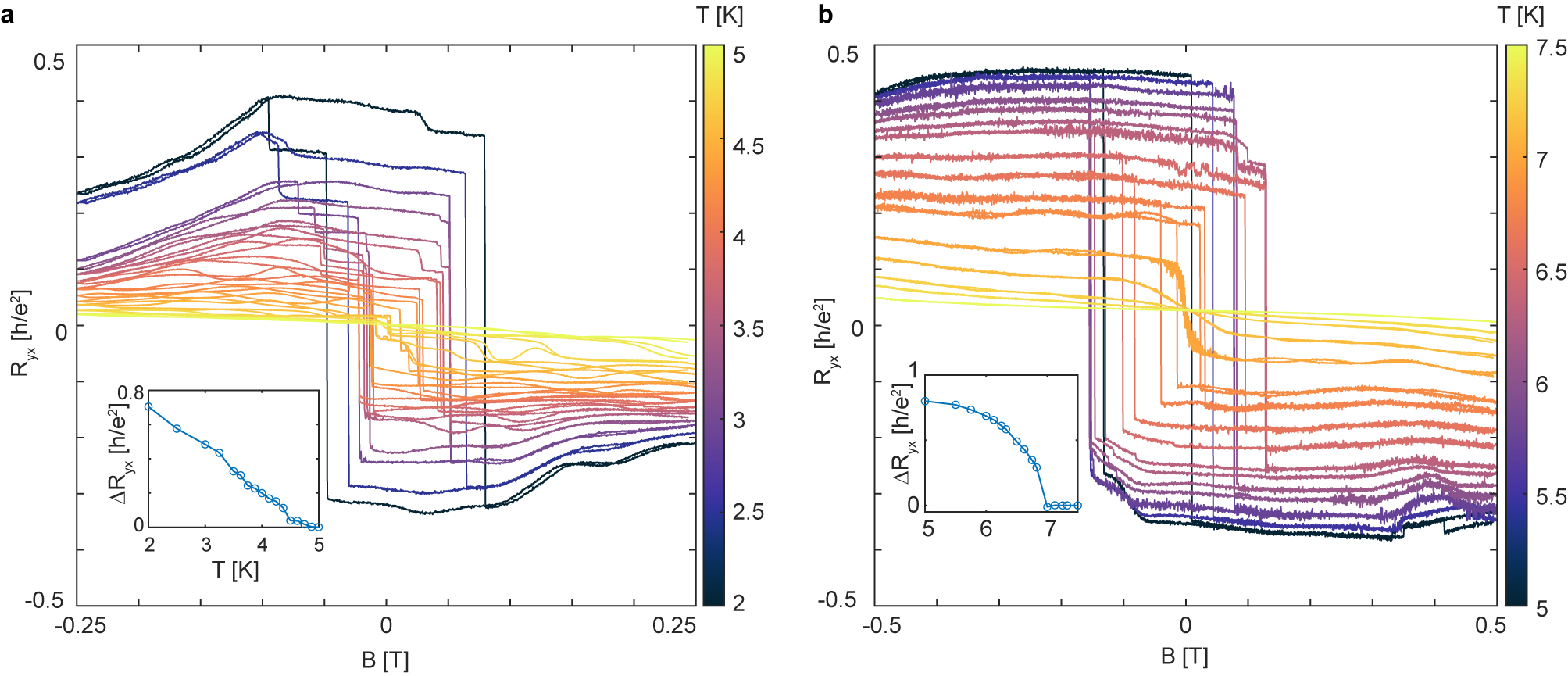} 
\caption{\textbf{Temperature dependence of the hysteresis at $\nu=1$ and 3 in device D1}.
\textbf{a, b}.
Insets show the temperature dependence of the height of the hysteresis loop height, as defined in Fig \ref{fig:3}a. 
}
\label{fig:S:tdephyst}
\end{figure*}

\begin{figure*}[ht!]
\includegraphics[]{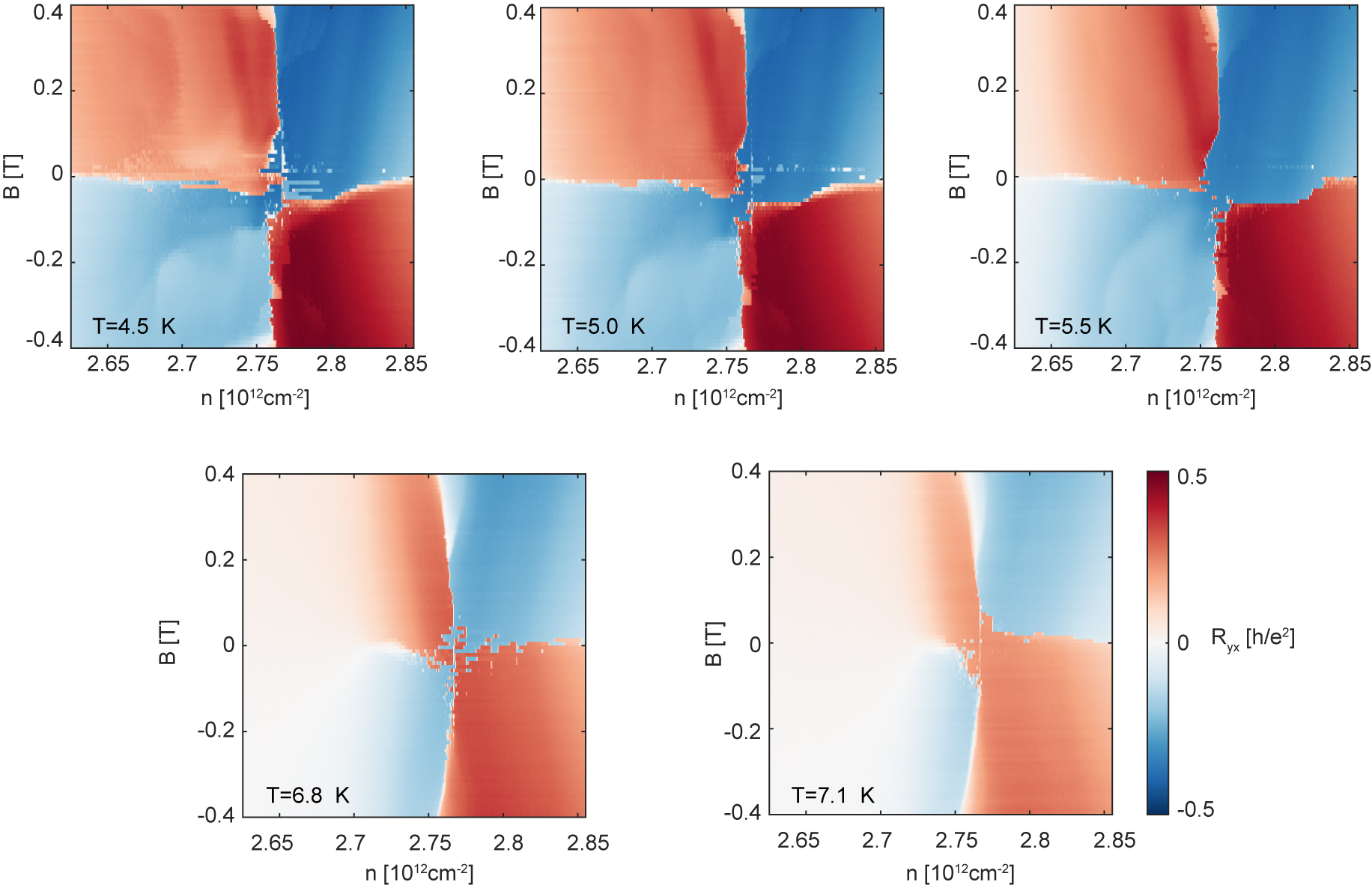} 
\caption{\textbf{$\mathbf{n}$ and $\mathbf{B}$ dependence of the measured anomalous Hall effect, plotted at selected temperatures for $\mathbf{D =0.4}$~V/nm in device D1.
}
Temperatures are labeled on individual panels.
\label{fig:S:tdepah}}
\end{figure*}

\begin{figure*}[ht!]
\includegraphics[]{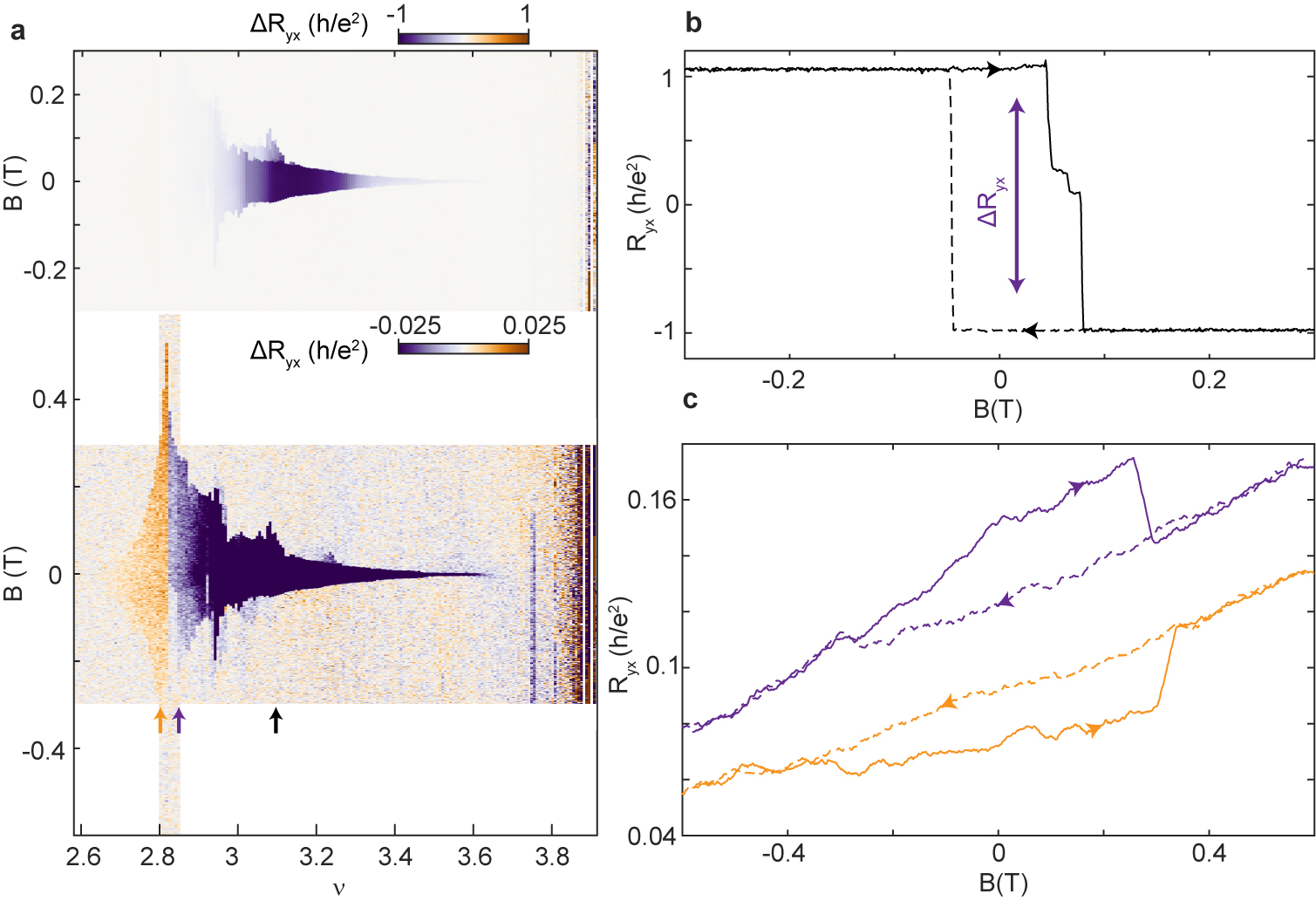} 
\caption{\textbf{Magnetization reversal in twisted bilayer graphene.}
\textbf{a}. Anomalous Hall resistance $\Delta R_{yx}$ associated with twisted bilayer graphene ferromagnetism, extracted by subtracting $R_{yx}(B)$ as $B$ is increased from $R_{yx}(B)$ as $B$ is decreased. The tBLG device is the same as in Ref.~\cite{serlin_intrinsic_2020}.  Colorscale is fixed to the von Klitzing constant in the top of the plot to show the range of filling factors for which a robust QAH effect is observed.  The colorscale axis is dramatically reduced in the bottom plot to illustrate weak features in $\Delta R_{yx}(\nu)$.  For $\nu < 3$, the coercive field of the ferromagnetic order increases dramatically, peaking at $\nu = 2.82$ electrons per moir\'e unit cell. For $\nu < 2.82$, $\Delta R_{yx}$ switches sign, indicating that the valley polarization of the ground state of the system at finite magnetic field has switched.  \textbf{b}. Robust Chern 1 QAH effect at $\nu = 3.1$. \textbf{c}. Ferromagnetic hysteresis plots on opposite sides of the divergence of the coercive field close to $\nu = 2.82$ (with offset).  Note the change in the relative sign of $\Delta R_{yx}$.  
}
\label{fig:s:tblg}
\end{figure*}

\begin{figure*}[ht!]
\includegraphics[]{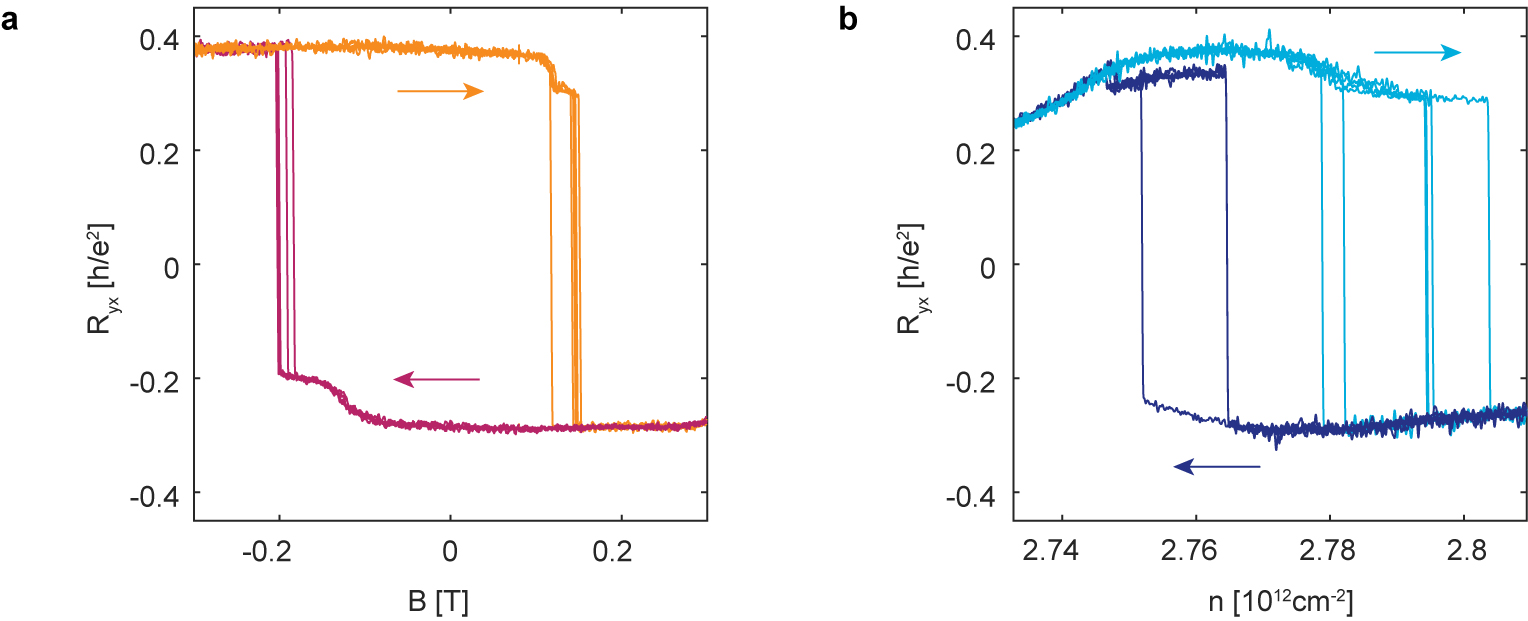} 
\caption{\textbf{Repeatability of magnetization switching with $\mathbf{B}$ and $\mathbf{n}$ measured in device D1}.
\textbf{a}. Repeated magnetic field hysteresis loops.
\textbf{b}. Repeated doping hysteresis loops.
Both panels taken under conditions analagous to those in Figure \ref{fig:4}b, described in the main text. 
}
\label{fig:S:hystrepeat}
\end{figure*}

\newpage\clearpage
\widetext
\begin{center}
\textbf{\large Supplementary Information}
\end{center}
\renewcommand{\figurename}{\textbf{Fig.}}
\renewcommand{\thefigure}{S\arabic{figure}}
\renewcommand{\thesubsection}{S\arabic{subsection}}
\setcounter{secnumdepth}{2}
\renewcommand{\theequation}{S\arabic{equation}}
\renewcommand{\thetable}{S\arabic{table}}
\setcounter{figure}{0}
\setcounter{equation}{0}
\onecolumngrid
\section{Moir\'e miniband simulation}\label{Sec_band_structure}
 The valley- and spin-projected $\pmb{k}$-space continuum model Hamiltonian of tMBG acts on
 six-component $\pmb{k} \cdot \pmb{p}$ spinors $\Psi = (\chi_{\sub{1A}}, \chi_{\sub{1B}}, \chi_{\sub{2A}}, \chi_{\sub{2B}}, \chi_{\sub{3A}}, \chi_{\sub{3B}})^{\sub{T}}$
 that describe slow spatial variations of the carbon $\pi$-orbital amplitudes on each of the trilayer's six sublattices.  
 Using a momentum-space representation for the spinor components,
 the Hamiltonian is 
\begin{equation}\label{Eq.hamil}
    H(\pmb{k})=
    \begin{pmatrix}
    h_{\theta/2}(\pmb{k}) & T & 0 \\
    T^\dagger & h_{-\theta/2}(\pmb{k}) & T_{\sub{Bernal}} \\
    0 & T_{\sub{Bernal}}^\dagger & h_{-\theta/2}(\pmb{k})
    \end{pmatrix}.
\end{equation}
Here $T$ is the interlayer tunneling between the monolayer and the adjacent Bernal layer\cite{bistritzer_moire_2011-1},
\begin{equation}
T_{\pmb{k'},\pmb{k}} = w \sum\limits_{j=1}^3 \delta_{\pmb{k'},\pmb{k}-\pmb{q}_j} T_j,
\end{equation}
where the sublattice-dependent hopping matrices
$T_j$ and the momentum jumps $\pmb{q}_j$ are defined in Ref.~\onlinecite{bistritzer_moire_2011-1}.
In our calculations we took the tunneling strength to be $w=117$ meV and 
reduced the diagonal elements of $T_j$ to $w_{\sub{AA}}=0.7 w_{\sub{AB}}=82$ meV 
to account\cite{Yoo_atomic_2019} for corrugation and strain effects. 
For the intralayer Hamiltonains we used $\pi$-band Dirac models\cite{neto_electronic_2009} 
that allow for sublattice-dependent energies 
$\varepsilon_{\sub{A}}$ and $\varepsilon_{\sub{B}}$:
\begin{equation}
    h_\theta(\pmb{k}) = 
    \begin{pmatrix}
        \varepsilon_{\sub{A}} & -\gamma_0 f(\pmb{k}) \\
        -\gamma_0 f^*(\pmb{k}) & \varepsilon_{\sub{B}}
    \end{pmatrix}
    \approx
     \begin{pmatrix}
         \varepsilon_{\sub{A}} & \hbar v_{\sub{D}} |\pmb{q}| e^{-i(\theta_{\pmb{q}} - \theta)} \\
         \hbar v_{\sub{D}} |\pmb{q}|  e^{i(\theta_{\pmb{q}} - \theta)} & \varepsilon_{\sub{B}}
     \end{pmatrix},
\end{equation}
where
\begin{equation}
f(\pmb{k})= \sum\limits_{j=1}^3 e^{i\pmb{k} \cdot \pmb{\delta}_j},
\end{equation}
$\pmb{\delta}_j$ are three honeycomb lattice nearest neighbour vectors,
and $\pmb{q} = \pmb{k} - \pmb{K}$ is the crystal momentum measured from 
graphene Brillouin-zone corner. We adopt the intralayer nearest neighbour hopping amplitude $\gamma_0=2610$ meV \textit{ab initio} calculated in Bernal-stacked bilayer graphene\cite{jung_accurate_2014}. The corresponding Dirac velocity $v_{\sub{D}} \approx 0.85 \times 10^6$ m/s is smaller than the commonly used value $10^6$ m/s in twisted bilayer graphene. It should be noted that Dirac velocity in the effective continuum model in Eq.(\ref{Eq.hamil}) would modify the band structure, larger $v_{\sub{D}}$ increases the single-particle band gap at charge neutrality.
Finally, $T_{\sub{Bernal}}$ is the interlayer tunneling between the 
middle and bottom Bernal-stacked layers:
\begin{equation}
    T_{\text{Bernal}} = 
    \begin{pmatrix}
        t_4f(\mathbf{k}) & t_3 f^*(\mathbf{k}) \\
        t_1 & t_4 f(\mathbf{k})
    \end{pmatrix}
\end{equation}
Here we define the interlayer hopping parameters as
\begin{equation}
\begin{split}
    t_1 &= \langle \mathbf{R}_{\sub{2B}}| \mathcal{H} | \mathbf{R}_{\sub{3A}} \rangle \\
    t_3 &= \langle \mathbf{R}_{\sub{2A}}| \mathcal{H} | \mathbf{R}_{\sub{3B}} \rangle \\
    t_4 &= \langle \mathbf{R}_{\sub{2A}}| \mathcal{H} | \mathbf{R}_{\sub{3A}} \rangle = \langle \mathbf{R}_{\sub{2B}}| \mathcal{H} | \mathbf{R}_{\sub{3B}} \rangle \\ \\
\end{split}
\end{equation}
$|\mathbf{R}_{\alpha} \rangle$ are $\pi$-orbitals. Note that some of these definitions are different in signs with the widespreadly used Slonczewski-Weiss-McClure model parameters\cite{mccann_electronic_2013}. We take $t_1=361$ meV, $t_3=283$ meV and $t_4=138$ meV\cite{jung_accurate_2014} in our calculations.  The sublattice energies were taken
to be identical on both sublattices within each graphene layer, thereby 
ignoring the possible influence of 
the hBN encapsulating, but layer-dependent on-site energies are considered to account 
for displacement fields.
The potential energy difference between layers was taken to be 
$\Delta_U = eDd/\epsilon_{\text{bg}}$ with $D$ the displacement field,
$d=3.3$~\AA \ the graphene interlayer separation, and $\epsilon_{\text{bg}}=4$
a background dielectric constant that accounts for remote-band
polarizations in graphene sheets.

\section{Self-consistent Hartree approximation}\label{Sec_HA}
Self-consistent Hartree approximation is extensively used to capture the band structure at finite carrier densities. We ignore the exchange interaction between carriers even though it was shown\cite{xie_nature_2020} to be important in magic angle twisted bilayer graphene.

In a dual-gated system, the external electric field and carrier density can be tuned individually. Assume a positive external displacement field $D$ is in the direction as depicted in Fig.1b (monolayer to bilayer) in the main text, onsite energies on top (monolayer), middle and bottom layers are respectively $U_1=-eDd/\epsilon_{\text{bg}}$, $U_2=0$ and $U_3=eDd/\epsilon_{\text{bg}}$. For a fixed total electron density $n_{\text{tot}} = n_1 + n_2 + n_3$, after charge redistribution the induced potential energies are
\begin{equation}
\begin{split}
    U_1^{\text{ind}} &= \frac{e^2d}{2\epsilon_{\text{bg}}} (n_1 - n_2 - n_3) \\
    U_2^{\text{ind}} &= 0 \\
    U_3^{\text{ind}} &= \frac{e^2d}{2\epsilon_{\text{bg}}} (-n_1 - n_2 + n_3) 
\end{split}
\end{equation}
where $n_1$, $n_2$ and $n_3$ are electron densities on each layer and are calculated by summing over wavefunction squares
\begin{equation}
    n_l = 4\sum\limits_{n,\pmb{k},l, \alpha} |\psi_{nl\alpha}(\pmb{k})|^2 f(\mu-\varepsilon_{n \pmb{k}}) - 4n_0
\end{equation}
$\pmb{k} \in$ moir\'e Brillouin zone, $n$, $l$ and $\alpha$ are band, layer and sublattice indices. Chemical potential $\mu$ is determined by total electron density. $n_0$ is the background density from negative Fermi seas. The integer 4 takes into account four flavors (two valleys and two spins) as we use a valley- and spin-polarized continuum Hamiltonian in Eq.~(\ref{Eq.hamil}).

\section{Electrical Reversibility}
\begin{figure*}[ht!]
\includegraphics[width=0.7\textwidth]{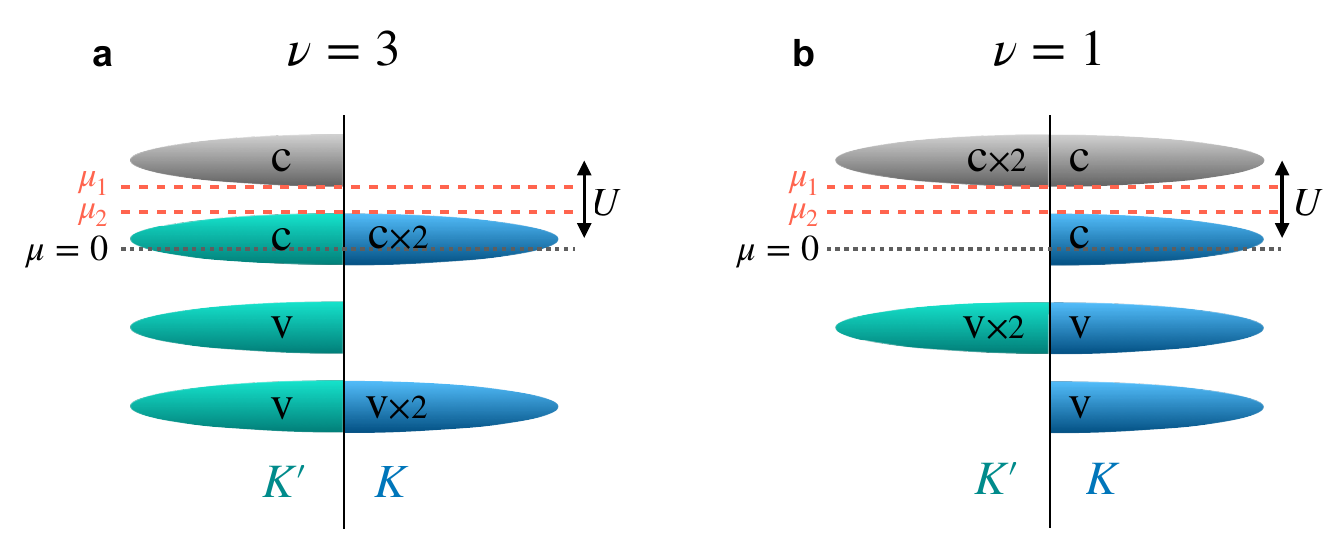} 
\caption{\textbf{Schematic Correlated Bands of Orbital Chern Insulator in Twisted Multilayer Graphene.}
For each flavor all band energies are shifted down by a constant exchange energy $U$ 
when the flat conduction band(s) is(are) occupied. 
\textbf{a} Fermi level interval $(\mu_2,\mu_1)$ in the $\nu=3$ gap.
\textbf{b} Fermi level interval $(\mu_2,\mu_1)$ in the $\nu=1$ gap.
}
\label{fig:S:bands}
\end{figure*}
Orbital Chern insulators have been observed in 
both tMBG with a finite displacement field and in magic-angle twisted
bilayer graphene (tBLG) encapsulated by hexagonal boron nitride  (hBN).
We note that the magnetization of an orbital Chern insulator can in 
principle change sign for a given sense of valley polarization 
at any band filling factor $\nu$, but that it is overwhelmingly more likely to change sign 
at integer $\nu$ where the magnetization can be discontinuous. 
We therefore focus on the issue of when the 
magnetization is likely to change sign across a gap characterized by a non-zero Chern number.
The magnetization of tBLG orbital Chern insulators 
has been previously discussed using a continuum model approach\cite{zhu_curious_2020}.  
Here we use
the same approach to compare theoretical expectations for magnetizations on opposite sides of the $\nu=1$ and $\nu=3$ gaps for both tBLG and tMBG.

As explained in\cite{zhu_curious_2020} the continuum model magnetization is a sum over separate contributions from
different spin and valley flavors.  We will account for time-reversal-symmetry breaking using a simplified 
mean-field approach in which the quasiparticle energies for a particular flavor 
are shifted downward by a momentum-independent exchange $U$ when the lowest energy
conduction band is occupied for that flavor, leaving the quasiparticle wavefunctions 
unchanged.  This prescription approximates the predictions of self-consistent 
Hartree-Fock mean-field theory\cite{xie_nature_2020}.
The resulting mean-field bands are illustrated schematically in Fig.~\ref{fig:S:bands}.  
Insulating states occur when the 
energy shift $U$ is greater than the bandwidth $w_\text{c}$ of the conduction band, resulting in a gap 
\begin{equation}
\Delta = \text{min}\{U-w_\text{c}, E_g\}.
\end{equation}
Here $E_g$ is the gap between the valence and conduction bands in the absence of the exchange interaction shift $U$.  
If we choose the zero of energy at the 
middle of $E_g$, it follows that for both $\nu=3$ and $\nu=1$,
the bottom of the conduction band of the unoccupied flavor (or flavors) is at energy $\mu_1=E_g/2$.  
The energy at the top of the highest occupied band is $\mu_2=\mu_1-\Delta$.

The requirement for a magnetization sign change across the gap can be expressed in terms of the magnetizations 
at chemical potential $\mu_1$ ($M(\mu_1)$) and at chemical potential $\mu_2$ ($M({\mu_2})$):
\begin{equation}
    M(\mu_1) \cdot M(\mu_2) < 0.
\end{equation}
$M({\mu_1})$ and $M({\mu_2})$ differ by the magnetization jump across the gap, $\delta M = M({\mu_1}) - M({\mu_2})$, 
which is proportional to the Chern number of the time-reversal partner of the unoccupied bands.  Assuming higher 
occupation of valley $K$, the case shown in Fig.~\ref{fig:S:bands},
\begin{equation}
    \delta M =  C_{\sub{cK}} \Delta/2\pi.
\end{equation}
where $C_{\sub{cK}}$ is the Chern number of the flat conduction band in valley $K$. 
Magnetization sign reversal will occur if $\delta M$ is 
large enough and has the correct sign:
\begin{equation}\label{criteria}
\begin{split}
    \delta M \cdot M_{\sub{bulk}} < 0 \\
    |\delta M| - |M_{\sub{bulk}}| > 0
\end{split}
\end{equation}
Here $M_{\sub{bulk}}$ is the total magnetization when the edge states 
in the gap are unoccupied, {\it i.e.} when the chemical potential is at the bottom of the gap:  $M_{\sub{bulk}}= M(\mu_2)$.

In order to identify the influence of the flavor-dependent energy shifts and 
occupations in the 
broken time-reversal symmetry state, we write the magnetization in the following form ($e=\hbar=1$): 
\begin{equation}
M(\mu) = \sum\limits_{m,f} \Big( \tilde{M}_{mf} n_{mf} + \frac{\mu C_{mf}n_{mf}}{2\pi} \Big) + U\sum\limits_{m,f \in f_{\text{shift}}} \frac{C_{mf} n_{mf}}{2\pi}.
\label{eq:M}
\end{equation}
Here $m$ is a band index, $f$ is a spin/valley flavor index,
$f_{\sub{shift}}$ is the set of flavors that have had energies shifted by conduction band occupation,
$n_{mf}$ is the band occupation, $C_{mf}$ is the band Chern number, and $\tilde{M}_{mf}$ is the
$M^1$ portion of the 
magnetization defined in Ref.~\cite{zhu_curious_2020}, evaluated with the 
zero of energy located at the middle of the gap $E_g$ for that flavor. 
In Eq.~(\ref{eq:M}) we have used that the
magnetization contribution of an occupied band shifts by 
$- C_{mf} \delta E/(2\pi)$ when band energies are rigidly shifted by $\delta E$.  

When we apply Eq.~(\ref{eq:M}) to $\nu=3$ we have two occupied conduction bands in valley $K$ and one
occupied conduction band in valley $K'$, as shown in Fig.~\ref{fig:S:bands}a.  
Since the band occupation numbers of all valence bands are 
equal and $\tilde{M}_{mK}=-\tilde{M}_{mK'}$ it follows that
\begin{equation}
\sum_{mf} \tilde{M}_{mf} n_{mf} = \tilde{M}_{\sub{cK}},  
\end{equation}
where $\tilde{M}_{\sub{cK}}$ is the $M^1$ part magnetization of the lowest energy conduction band in valley $K$. 
Since opposite valleys also have opposite Chern numbers, the sum of the Chern numbers over all occupied bands comes only from the 
uncompensated conduction bands and is $C_{\sub{cK}}$.  
On the other hand the sum of the Chern numbers over all occupied bands that 
suffer an energy shift is
$C_{\sub{cK}}+C_{\sub{vK}}+C_{\sub{v'K}}$
where $C_{\sub{vK}}$ is the Chern number of the highest energy valence band in valley $K$
and $C_{\sub{v'K}}$ is the corresponding sum over all remote valence bands. 
The final expression for the magnetization as a function of chemical potential in the $\nu=3$ gap is 
\begin{equation} 
M(\mu) = \tilde{M}_{\sub{cK}} + \frac{U (C_{\sub{cK}}+C_{\sub{vK}}+C_{\sub{v'K}}) + \mu C_{\sub{cK}}}{2\pi}.
\end{equation}
The same equation applies for $\nu=1$ since the two conditions differ by 
changing the occupation numbers and shifting the energies of two bands
that are in opposite valleys.  We emphasize that the energy shift contribution is 
proportional to the sum of the Chern numbers of all
shifted bands, and has a contribution from the valence bands because 
the energy shifts are valley-dependent.
On the other hand, the chemical potential shift contribution is proportional 
to the sum of all occupied-band Chern numbers, and this
does not have a valence band contribution. 

\begin{figure*}[ht!]
\includegraphics[width=\textwidth]{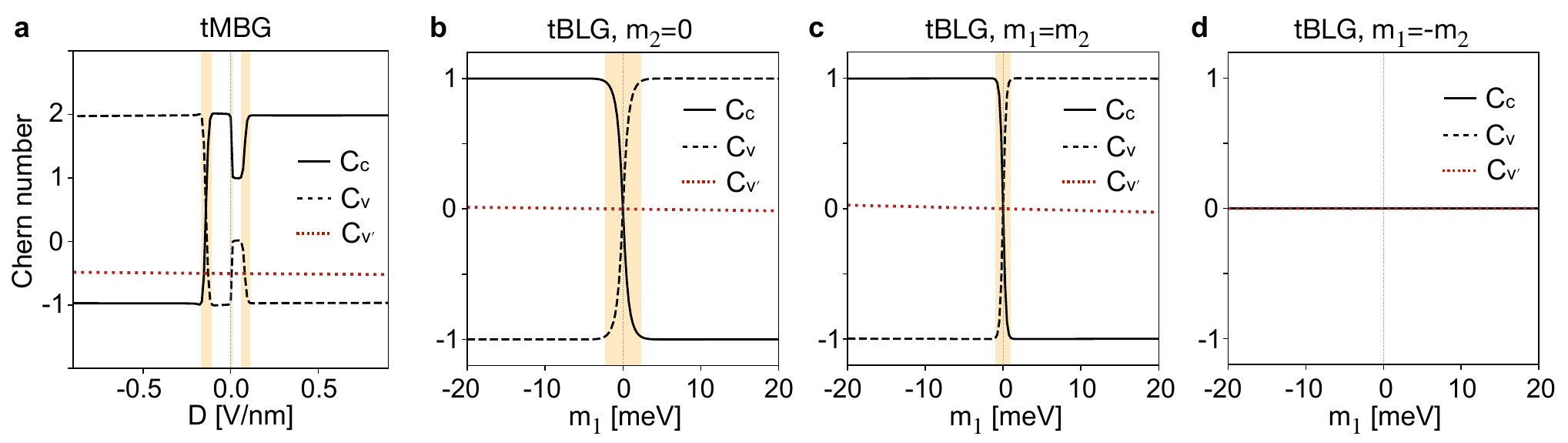} 
\caption{\textbf{Valley $K$ Chern numbers.}
$C_\text{c}$, $C_\text{v}$ and $C_{\text{v}'}$ are Chern numbers of the flat conduction band, flat valence band and integrated Berry curvatures summed over all remote valence bands.
\textbf{a} Chern numbers of $1.25^\circ$ tMBG {\it vs.} displacement field $D$. 
The two flat bands touch, either at $\kappa$ or $\kappa'$, near the yellow-shaded region.
\textbf{b-d} Chern numbers of $1.1^\circ$ tBLG.
\textbf{b} One-side alignment model {\it vs.} $m_1$ where the mass of the other graphene layer $m_2=0$.
\textbf{c} Two-side alignment model {\it vs.} $m_1=m_2$.
\textbf{d} Two-side alignment model {\it vs.} $m_1=-m_2$.
}
\label{fig:S:chern}
\end{figure*}

In Fig.~\ref{fig:S:chern}a, we plot Chern numbers
$C_{\sub{cK}}$, $C_{\sub{vK}}$ and $C_{\sub{v'K}}$ of $1.25^\circ$ tMBG {\it vs.} displacement field $D$.
$C_{\sub{cK}}+C_{\sub{vK}}=1$ and $C_{\sub{v'K}}=-0.5$ are both independent of $D$.
The Chern numbers are obtained by momentum space integration.
The two flat bands touch either at $\kappa$ or $\kappa'$ near the yellow-shaded regions 
(where the numerical results become inaccurate), transferring Berry curvature between them. 
The half-odd-integer value of $C_{\sub{v'K}}$ is expected in a continuum model with 
an odd number of graphene layers since the Berry curvature integrated over all valence band states
is $\pm 3/2$ in the decoupled trilayer limit when sublattice symmetry is weakly broken.  
When moir\'e bands are formed,
the integrated Berry curvature of each isolated moir\'e miniband
must however be an integer.  In gate biased tMBG our results show that 
integers are achieved by transferring Berry curvature between conduction and valence 
bands with a sense that depends on the sign of $D$.  
Note that both $M(\mu_1)$ and $M(\mu_2)$, but not their 
difference, is dependent on $C_{\sub{v'K}}$.  
This feature of our magnetization calculations is the continuum model 
manifestation of the band-Hamiltonian property\cite{raoux_orbital_2015} that the orbital magnetization
{\it vs.} $\mu$ curve over some narrow interval can be shifted by a constant by band rearrangements at very remote energies.
Our continuum model estimates are therefore more uncertain for the mean value of $M(\mu_1)$ and $M(\mu_2)$ 
than they are for their difference $\delta M$.

\begin{figure*}[ht!]
\includegraphics[width=\textwidth]{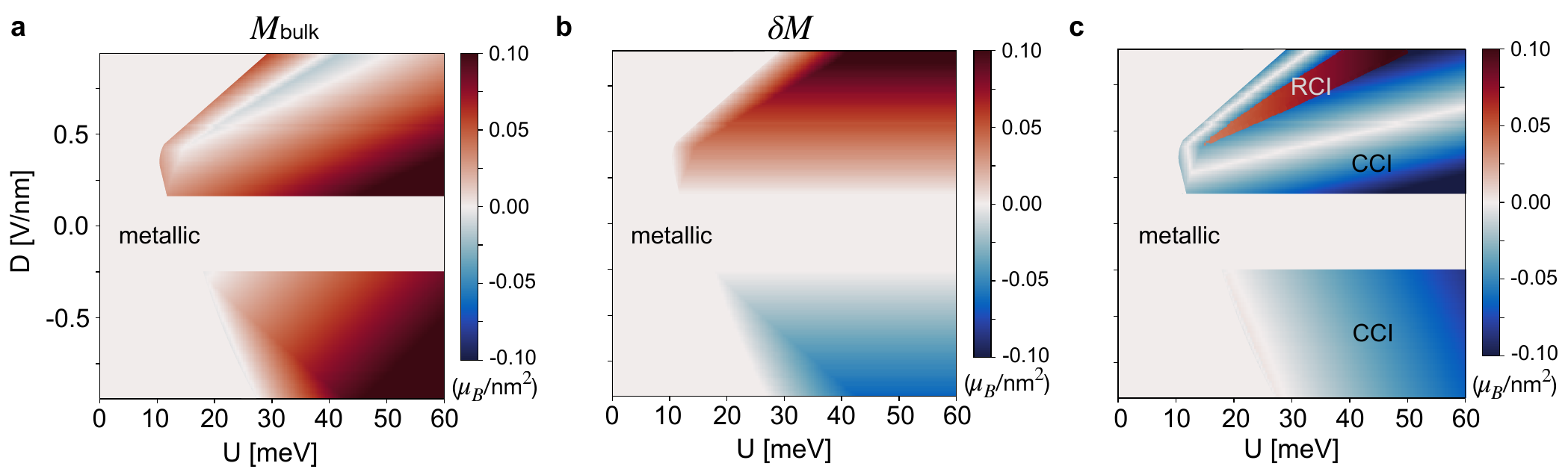} 
\caption{\textbf{Phase diagram {\it vs.} $D$ and exchange energy $U$ at $\nu=3$ for $1.25^\circ$ tMBG.}
\textbf{a} $M_{\sub{bulk}}$.
\textbf{b}
$\delta M$.
\textbf{c} The magnetude is the absolute value of $|\delta M| - |M_{\sub{bulk}}|$. The positive (negative) sign represents Eq.(\ref{criteria}) is satisfied (unsatisfied). The phase diagram identifies regions where the state is insulating, $M_\text{bulk}$ and $\delta M$ are opposite in sign and $|\delta M| - |M_\text{bulk}| > 0$ as reversible Chern insulator (RCI), and other regions where the state is insulating as conventional Chern insulator (CCI).
}
\label{fig:S:PhaseDiag_nu3}
\end{figure*}

As a comparison, we also show the corresponding Chern numbers (Fig.~\ref{fig:S:chern}b-d) of $1.1^\circ$ tBLG. The driver of non-trivial topology in the valley-projected bands in this case is sublattice symmetry breaking by 
encapsulating hBN layers (or spontaneously due to electron electron interactions) which induces masses (non-zero values of $m_l=(\varepsilon_{\sub{A}l}-\varepsilon_{\sub{B}l})/2$) in adjacent layers $l$. 
For one-side alignment case, {\it i.e.} only one graphene layer is nearly aligned with hBN, the mass on one graphene layer $m_1 \neq 0$ and the mass on the other graphene layer $m_2=0$. In this case, shown in Fig.~\ref{fig:S:chern}b, Chern numbers of flat bands are $C_{\text{c(v)}}=\pm 1$ and the total Chern number of all remote valence bands is $C_{\text{v}'}=0$. Moir\'e miniband formation no longer forces a transfer of Berry curvatures between conduction and valence bands, and none occurs. 
For two-side alignment case,
when $m_1=m_2 \neq 0$, corresponding to both graphene layers being nearly aligned with the surrounding hBN layers and having the same relative orientation, the flat band 
Chern numbers shown in Fig.~\ref{fig:S:chern}c are again
$C_{\text{c(v)}}=\pm 1$, and the total Chern number of all remote valence bands is again $C_{\text{v}'}=0$. 
If both graphene layers are nearly aligned but have opposite relative orientations with hBN, i.e. $m_1=-m_2$, the Chern numbers vanish as shown in Fig.~\ref{fig:S:chern}d.

Given these results for the Chern numbers the following two equations apply at both 
$\nu=3$ and at $\nu=1$ in tMBG:
\begin{equation}
M(\mu_1) = \tilde{M}_{\sub{cK}} + \frac{E_{g}C_{\sub{cK}}/2+U/2}{2\pi},
\label{M1MBG}
\end{equation} 
and 
\begin{equation}
M_{\sub{bulk}} \equiv M(\mu_2) = \tilde{M}_{\sub{cK}} + \frac{(E_{g}/2-\Delta)C_{\sub{cK}}+U/2}{2\pi}.
\label{M2MBG}
\end{equation}
The corresponding equations for tBLG are:
\begin{equation}\label{eq:M1BG}
M(\mu_1) = \tilde{M}_{\sub{cK}} + \frac{E_{g}C_{\sub{cK}}/2}{2\pi},
\end{equation} 
and 
\begin{equation}
M_{\sub{bulk}} \equiv M(\mu_2) = \tilde{M}_{\sub{cK}} + \frac{(E_{g}/2 -  \Delta)C_{\sub{cK}} }{2\pi}.
\label{M2BG}
\end{equation}
In the tMBG case, polarization toward valley $K$ implies a positive conduction band Chern number 
$C_{\sub{cK}}$ for positive displacement field, and therefore implies that $\delta M >0$.
Electrical reversal is possible only if $M_{\sub{bulk}}$ is negative for positive displacement field
and the gap $\Delta$ is larger than $\pi |M_{\sub{bulk}}|$.  
For the tBLG case, polarization toward valley $K$ implies a positive (negative) conduction band Chern number for negative (positive) $m_1$.  
Electrical reversal therefore requires that 
$M_{\sub{bulk}}$ is negative (positive) for negative (positive) $m_1$
and that the gap $\Delta$ is larger than 
$2\pi |M_{\sub{bulk}}|$.  

Figures~\ref{fig:S:PhaseDiag_nu3}a-c show the dependence of  $M_{\sub{bulk}}$, $\delta M$, and $|\delta M|-|M_{\sub{bulk}}|$ in tMBG on $U$ and $D$ after self-consistent Hartree approximation.
It should be noted that these results are calculated using a phenomenologically increased band gap $E_g$, on the order of 10 meV, to match experimental band gap measurements, owing to the fact that Hartree approximation alone will underestimate the gap. As we mentioned above, $M_{\sub{bulk}}$ is not accurately captured in our continuum model even though it is essential in the sign-reversal effect.
Gapless regions that correspond to  ($U<w_\text{c}$) or overlapping conduction and valence bands are marked as "metallic" in Fig.~\ref{fig:S:PhaseDiag_nu3}.
Figure~\ref{fig:S:PhaseDiag_nu3}c identifies regions where $M_{\sub{bulk}}$ and $\delta M$ are opposite in sign 
and $\delta M$ has a larger magnitude ({\textit i.e.} Eq.~(\ref{criteria}) is satisfied) as reversible Chern insulators (RCI), and other regions where the state is insulating 
as conventional Chern insulators (CCI). 
We see that RCI states occur for 
positive displacement fields when the exchange energy $U$ is large enough.  
Since we expect the effective value of $U$ to be smaller when the filling factor is closer to 
zero and the overall flat band system is therefore closer to half-filling -- enhancing
screening and correlation corrections to mean field theory, the experimental finding that RCI 
state occurs for $\nu=3$ but not for $\nu=1$ is consistent with Fig.~\ref{fig:S:PhaseDiag_nu3}.
 It should be pointed out that we are not considering hBN alignments, which break sublattice symmetry in tMBG, in our numerical calculations. However some specific hBN alignments can make the magnetization sign-reversal effect more robust numerically.

\begin{figure*}[ht!]
\includegraphics[width=\textwidth]{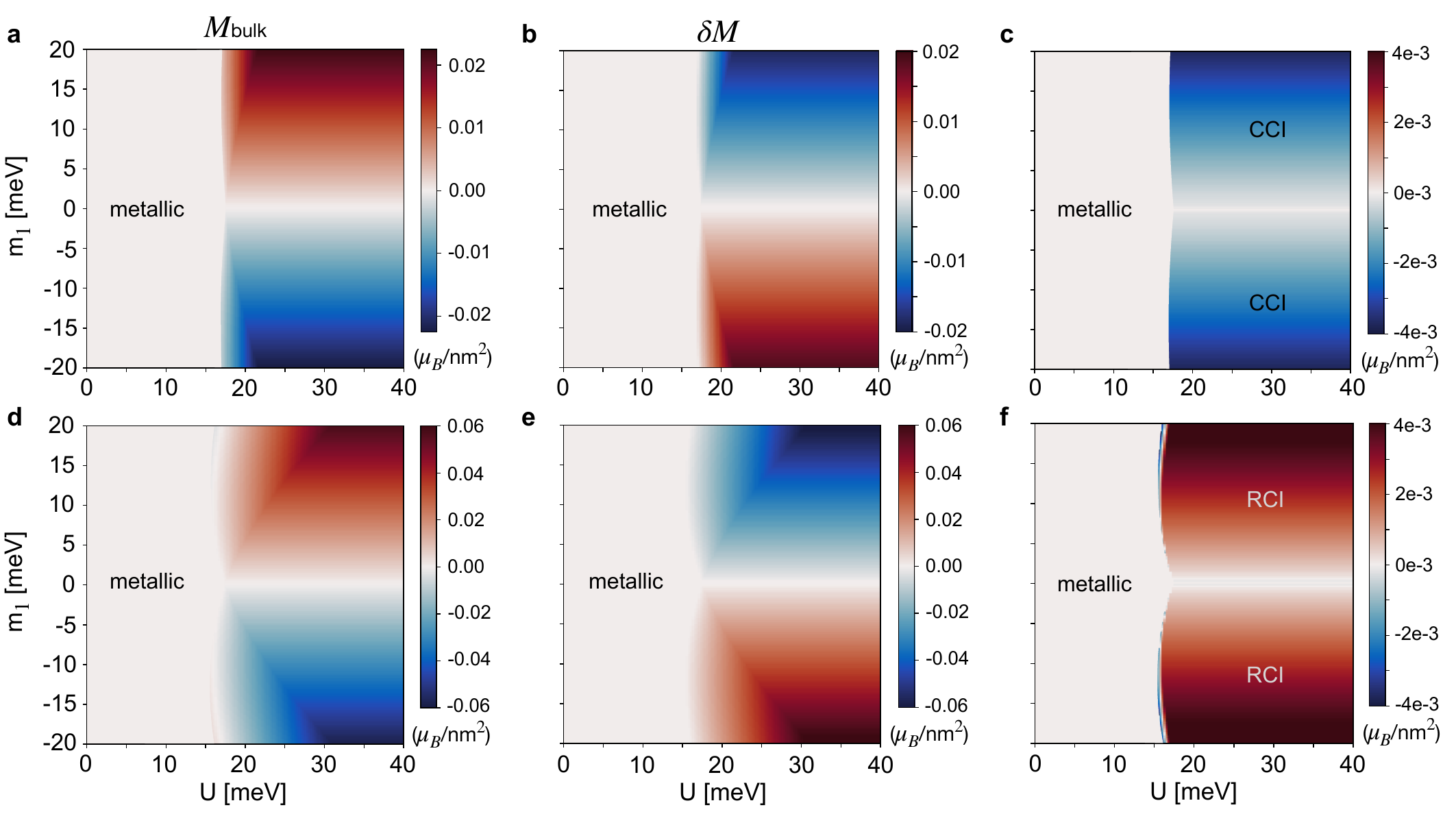} 
\caption{\textbf{Phase diagrams {\it vs.} $m_1$ and exchange energy $U$ at $\nu=3$ for $1.1^\circ$ tBLG.}
\textbf{a-c} One-side alignment model {\it vs.} $m_1$ and $U$.
\textbf{d-f} Two-side alignment model with $m_1=m_2$.
\textbf{a,d} $M_{\sub{bulk}}$.
\textbf{b,e} $\delta M$.
\textbf{c,f} The magnetude is the absolute value of $|\delta M| - |M_{\sub{bulk}}|$. The positive (negative) sign represents Eq.(\ref{criteria}) is satisfied (unsatisfied). In these model calculations without electron-electron interactions, gaps appear when $U$ is larger than 
$w_{\text{c}}$ and always results in Chern insulators, but these are electrically reversible only with two-sided alignment.
}
\label{fig:S:PhaseDiag_tBLG}
\end{figure*}

As a comparison, we also show the corresponding phase diagrams (Fig.~\ref{fig:S:PhaseDiag_tBLG}) of $1.1^\circ$ tBLG. 
Note that insulating states appear only beyond a minimum $U$ but require only infinitesimal masses.  Interestingly we find CCI states for 
one-side hBN alignment and RCI states for two-side hBN alignment. 
As we see in Fig.~\ref{fig:S:PhaseDiag_tBLG} the difference can be traced to
a difference in the signs of $M(\mu_1)$ in the two cases:
\begin{equation}
M(\mu_1) =
    \begin{cases}
         |M_{\sub{bulk}}| - |\delta M|, \text{  if $m_1>0$} \\
        |\delta M| - |M_{\sub{bulk}}|, \text{  if $m_1<0$} \\
    \end{cases}
\end{equation}
The sign of $M(\mu_1)$ is dependent on a competition between the $\tilde{M}_{\sub{cK}}$ and the
Chern number term in Eq.~(\ref{eq:M1BG}).  Two-sided alignment increases 
sublattice-symmetry breaking and increases both the magnitudes of the
conduction-valence energy-gap $E_{g}$ and $\tilde{M}_{\sub{cK}}$.  In our numerical calculations, however, the magnitudes of $\tilde{M}_{\sub{cK}}$ is doubled by two-side alignment, whereas the 
energy gap $E_g$ is increased by more than a factor of three, changing the sign of 
$M(\mu_1)$.  Since we expect that electron-electron interactions will also enhance
the energy gap by a larger factor than they enhance $\tilde{M}_{\sub{cK}}$, 
RCI behavior likely occurs for one-side alignment as well
when interaction effects are described in greater detail.  

\end{document}